\documentclass[11pt,titlepage]{article}

\usepackage{comment}
\usepackage{color}
\usepackage{setspace}
\usepackage{graphicx}
\usepackage{epstopdf}
\usepackage{bbm}
\usepackage{rotating}
\usepackage{textcomp}

\usepackage{amsfonts}
\usepackage{amssymb}
\usepackage{amsmath}
\usepackage{graphicx}
\usepackage{dsfont}
\usepackage{epsfig}
\usepackage{ae}
\usepackage{hyperref}
\usepackage{harvard}
\usepackage{amsmath}

\newtheorem{Prop}{Proposition}
\newtheorem{Cor}{Corollary}

\newenvironment{proof}[1][Proof]{\noindent\textbf{#1.} }{\ \rule{0.5em}{0.5em}}
\newtheorem{theo}{Theorem}[section]
\newtheorem{pr}{Proposition}[section]
\newtheorem{co}{Corollary}[section]

\newtheorem{exa}{Example}[section]
\newtheorem{lem}{Lemma}[section]
\newtheorem{rem}{Remark}[section]

\def\proof{\noindent \textbf{Proof:}\, }

\def\be{\begin{equation}} 
\def\ee{\end{equation}} 
\def\beqn{\begin{eqnarray}} 
\def\eeqn{\end{eqnarray}} 
\def\beq{\begin{eqnarray*}} 
\def\eeq{\end{eqnarray*}} 
\def\ba{\begin{array}} 
\def\ea{\end{array}} 
\newcommand{\bt}{\begin{theo}}
\newcommand{\et}{\end{theo}}
\newcommand{\bl}{\begin{lem}}
\newcommand{\el}{\end{lem}}
\newcommand{\bc}{\begin{co}}
\newcommand{\ec}{\end{co}}
\newcommand{\bp}{\begin{pr}}
\newcommand{\ep}{\end{pr}}
\newcommand{\bex}{\begin{exa}}
\newcommand{\eex}{\end{exa}\vspace{-4mm}}
\newcommand{\br}{\begin{re}}
\newcommand{\er}{\end{re}\vspace{-3mm}}

\definecolor{myBlue}{rgb}{0.80,0.85,1.00}
\definecolor{myYellow}{rgb}{0,1.000,0}

\newbox\treebox
\def\tree{\global\setbox\treebox=\boxtree}

\def\endsubtree{\ettext \egroup}

\newif\iftreetext\treetextfalse
\def\boxtree{\hbox\bgroup
  \baselineskip 2.5ex
  \tabskip 0pt
  \vbox\bgroup
  \treetexttrue
  \let\par\crcr \obeylines
  \halign\bgroup##\hfil\cr}
\def\ettext{\iftreetext
  \crcr\egroup \egroup \fi}

\def\cons#1#2{\edef#2{\xmark #1#2}}
\def\car#1{\expandafter\docar#1\docar}
\def\docar\xmark#1\xmark#2\docar{#1}
\def\cdr#1{\expandafter\docdr#1\docdr#1}
\def\docdr\xmark#1\xmark#2\docdr#3{\def#3{\xmark #2}}
\def\xmark{\noexpand\xmark}
\def\nil{\xmark}

\def\settreesizes{\setbox0=\copy\treebox \global\let\treesizes\nil \setsizes}
\newdimen\treewidth
\def\setsizes{\setbox0=\hbox\bgroup
  \unhbox0\unskip
  \inittreewidth
  \sizesubtrees
  \sizelevel
  \egroup}

\def\inittreewidth{\ifx\treesizes\nil   
  \treewidth=0pt                        
\else \treewidth=\car\treesizes         
  \global\cdr\treesizes                 
  \fi}                                  

\def\sizesubtrees{\loop                 
  \setbox0=\lastbox \unskip             
  \ifhbox0 \setsizes                    
  \repeat}                              

\def\sizelevel{\ifdim\treewidth<\wd0    
  \treewidth=\wd0 \fi                   
\global\cons{\the\treewidth}\treesizes} 

\newdimen\treeheight                    
\newif\ifleaf                           
\newif\ifbotsub                         
\newif\iftopsub                         
\def\maketree{\hbox{\treewidth=\car\treesizes  
  \cdr\treesizes                        
  \makesubtreebox\unskip                
  \ifleaf \makeleaf                     
  \else \makeparent \fi}}               

{\catcode`@=11                          
\gdef\makesubtreebox{\unhbox\treebox    
  \unskip\global\setbox\treebox\lastbox 
  \ifvbox\treebox                       
    \global\leaftrue \let\next\relax    
  \else \botsubtrue                     
    \setbox0\box\voidb@x                
    \botsubtrue \let\next\makesubtree   
  \fi \next}}                           

\def\makesubtree{\setbox1\maketree      
  \unskip\global\setbox\treebox\lastbox 
  \treeheight=\ht1                      
  \advance\treeheight 2ex               
  \ifhbox\treebox \topsubfalse          
    \else \topsubtrue \fi               
  \addsubtreebox                        
  \iftopsub \global\leaffalse           
    \let\next\relax \else               
    \botsubfalse \let\next\makesubtree  
  \fi \next}                            

\def\addsubtreebox{\setbox0=\vbox{\subtreebox\unvbox0}}
\def\subtreebox{\hbox\bgroup            
  \vbox to \treeheight\bgroup           
   \ifbotsub \iftopsub \vfil            
       \hrule width 0.4pt               
     \else \treehalfrule \fi \vfil      
   \else \iftopsub \vfil \treehalfrule  
     \else \hrule width 0.4pt height \treeheight \fi\fi 
   \egroup                              
  \treectrbox{\hrule width 1em}\hskip 0.2em\treectrbox{\box1}\egroup}

\def\treectrbox#1{\vbox to \treeheight{\vfil #1\vfil}}
\def\treehalfrule{\dimen0=\treeheight   
  \divide\dimen0 2\advance\dimen0 0.2pt 
  \hrule width 0.4pt height \dimen0}    

\def\makeleaf{\box\treebox}             
\def\makeparent{\ifdim\ht\treebox>\ht0  
  \treeheight=\ht\treebox               
\else \treeheight=\ht0 \fi              
\advance\treewidth-\wd\treebox          
\advance\treewidth 1em                  
\treectrbox{\box\treebox}\hskip 0.2em   
\treectrbox{\hrule width \treewidth}\treectrbox{\box0}} 

\begin{document}

\pagestyle{plain}

 \nocite{*}
\bibliographystyle{econometrica}

\author{Weidong Tian\\
University of North Carolina at Charlotte 
\\
\and 
Zimu Zhu
\\ University of Southern California
 }
 \date{}
 
 \renewcommand\footnotemark{}
	\title{\Large \bf  A Portfolio Choice Problem Under Risk Capacity Constraint}
	\thanks{
Corresponding author: Weidong Tian, Belk College of Business, University of North Carolina at Charlotte. Email: wtian1@uncc.edu. Zimu Zhu, Department of Mathematics, University of Southern California. Email:zimuzhu@usc.edu. We thank Jianfeng Zhang and Tao Pang for stimulating discussions on this paper. We also thank Dr.Xiaojing Xing for her assistance in numerical implementation. The authors would like to thank the editor and anonymous referee for their constructive comments and suggestions.}

\maketitle
\thispagestyle{empty}
	\bigskip

	\clearpage
	
	
	
\renewcommand{\abstractname}{
{\LARGE A Portfolio Choice Problem Under Risk Capacity Constraint }\\[1in] Abstract}

\begin{abstract}
This paper studies the asset allocation problem for a retiree facing longevity risk and living standard risk. We introduce a risk capacity constraint to reduce the living standard risk in the retirement period. Whether the retiree focuses on intertemporal consumption or inheritance wealth, we demonstrate a unique number to measure the expected lump sum of the spending post-retirement. The optimal portfolio is nearly neutral to the stock market movement if the portfolio's value is higher than this finite critical value; otherwise, the retiree actively invests in the stock market. As a comparison, we consider a dynamic leverage constraint and show that the corresponding optimal portfolio would lose significantly in stressed markets.

 \vspace{0.9cm}



\end{abstract}

\pagebreak

\nopagebreak
\newpage

\setcounter{page}{1}

\parskip 0.5em

\renewcommand{\thefootnote}{\arabic{footnote}}


\section{Introduction}
Investment after retirement is significantly different from investment for (before) retirement in several respects. Retirees invest in an unknown while finite length of time because of longevity (mortality) risk. They also worry about the balance between spending and leaving wealth as an inheritance. More importantly, because of no labor income, these individuals will face living standard risk if a market downturn occurs, leading to a substantial decline in their living standards.\footnote{As Kenneth French presented at the Annual Conference for Dimensional Funds Advisors, 2016, ``It is living standard risk you should know about the risk. It is what your exposure is to a major change in your standard of living during the entirely uncertain numbers of years you remain alive.'' } 

Motivating by the asset allocation problem after retirement, this paper studies a portfolio choice problem with two distinguishing features. One is an uncertain investment time horizon; another is that the investment dollar amount in the risky asset is always bound from above by a fixed constant (capacity). We name this constraint a ``{\em risk capacity constraint}". We present the asset allocation problem after retirement as an optimal portfolio choice problem. Precisely, the retiree's mortality risk is formulated by an uncertain investment time horizon. The length of each individual's retirement may differ from the statistical life expectancy, and the mortality risk is virtually independent of the market risk in the financial market. Since the absence of labor income results in severe concerns about living standards, we use the risk capacity constraint to address the retiree's concern about the living standard risk after retirement.\footnote{Since we focus on the risk capacity, we ignore other factors such as health care risk and real estate assets in retirement portfolios. See, for instance, Yogo (2016) about the discussion of other factors.} 

We first characterize the value function of the optimal portfolio choice problem. Then we characterize the region of the wealth in which the risk capacity constraint is binding. Specifically, assuming the value function is $C^2$ smooth, and under some technical conditions about the value function, then there exists a threshold (a positive number $W^*$) such that the retiree invests the capacity amount in the risky asset if and only if the portfolio wealth is greater than the threshold. Otherwise, the investment amount is strictly smaller than the capacity. 

To derive the consumption-investment policy explicitly, we next investigate two particular yet critical situations in which these technical conditions of the value function can be verified. In the first situation, Problem (A), the investor (retiree) focuses on intertemporal consumption, whereas the investor concentrates on the inheritance wealth in the second situation, Problem (B). For a tractable purpose, we study the CRRA utility functions. In Problem (A) and Problem (B), if the value function is $C^2$ smooth, we demonstrate the existence of one number $W^*$ such that the risk capacity constraint is binding or not depends on whether the portfolio wealth is greater or smaller than this number. Moreover, we derive the explicit consumption-investment policy in terms of this threshold.
  

  For Problem (A), the $C^2$ smooth property of the value function follows from the classical study in Zariphopoulou (1994). In general, when the utility function satisfies a global Lipschitz condition, and the invested dollar amount is bounded from below and above by two positive numbers, the value function is $C^2$ smooth by a recent remarkable theorem of Strulovic and Szydlowski (2015). However, the CRRA utility does not satisfy the global Lipschitz condition, and the risk capacity constraint implies that the dollar amount can be any sufficiently small positive number. A technical component in this paper is to investigate the smooth property of the value function and thus the explicit expression of the value function in Problem (B). We reduce the $C^2$ smooth property of the value function to solve a nonlinear equation of one variable. Briefly speaking, there exists a positive real number solution (and  unique upon its existence) of the nonlinear equation if and only if the value function is $C^2$ smooth. Moreover, this real number solution separates the unconstrained region and constrained region of the risk capacity constraint. We also show that this method to determine the threshold $W^*$ and the value function explicitly can be used in Problem (A).

 Our solution to the optimal portfolio choice problem provides new theoretical insights for retirement investment.  
First, the optimal investment strategy displays a {\em wealth-cycle} property, in contrast to the {\em life-cycle} feature which is suggested in Modigliani (1986). That is, individual's investment and consumption decision has a life-cycle feature. See Bodie, Merton, and Samuelson (1992), Cocco, Gomes, and Maenhout (2005), Gomes and Michaelides (2005),  Benzoni, Dufresne, and Goldstein (2007), and  Bodie, Detemple, and Rindisbacher (2009) for life-cycle theoretical and empirical studies.\footnote{The life-cycle hypothesis is also used for preparing the retirement portfolio and in the retirement portfolio.  For instance, a conventional rule for an agent of age $t$ is to invest (100 - t)/100 percent of the wealth in the stock market.  See Malkiel (1999). Even after retirement, a time-dependent (life-cycle) investment strategy is also popular among financial advisors.}
 Specifically,  when the portfolio value is unsustainable for the entire retirement period, the retiree should invest in the market because the dollar investment in the stock will increase the expected portfolio value. Nevertheless, the percentage of wealth in the market {\em declines} in the wealth. The declining percentage of the wealth invested in risky assets is due to the retiree' living standard concern to protect the portfolio value. This decreasing feature of the percentage becomes significant when the portfolio's worth is sufficiently high.  
 
Second, since the dollar invested in the stock is always a constant $L$ when the portfolio wealth is higher than a threshold $W^*$, this threshold $W^*$ measures the expected lump sum of the spending in the retirement period. Intuitively, when the portfolio is worth more than this threshold, the retiree aims to protect the portfolio by investing only a fixed amount of $L$ in the stock market without losing the living standard. This {\em contingent constant-dollar strategy} is thus a buffer-invest strategy: when the wealth is below the target, the retired invests; and if the wealth is above the target, the retire will dis-invest more on the stock market. In a classical constant-dollar strategy, the dollar invested in the risky asset is always fixed. In contrast, by a contingent constant-dollar strategy in this paper we mean a fixed dollar is invested in the risky asset only when the portfolio value is higher than a threshold. It is also different from the constant proportion portfolio insurance strategy in Black and Perold (1992) and El Karoui and Jeanblanc-Picque (1998), which is interpreting by a put option on the Merton-type portfolio and the underlying Merton-type portfolio. See Carroll (1997) for a buffer-stock saving theory under the permanent income hypothesis. Our result is consistent with the retirement portfolio's {\bf decumulation} process.  In contrast, investing for (before) retirement is an {\em accumulating} asset process.
 
Third, the portfolio is nearly {\em independent} of the stock market when the retiree's portfolio is worth sufficiently embracing the living standard. Moreover, the portfolio risk is also nearly zero if the retiree's portfolio wealth is high enough. Therefore, the retiree's living standard risk is reduced substantially under the risk capacity constraint. 

Fourth, we demonstrate that the {\em sub-optimality} of the annuity, a popular insurance product for a retiree. Annuities are the perfect financial vehicle to counter only the mortality risk. However, due to unexpected costs or shock, non-annuitized wealth could be needed to cover the bill. From a portfolio choice perspective, we demonstrate that the annuity is not optimal since the optimal consumption should depend on wealth by incorporating a risk capacity constraint. The relationship between the consumption rate and wealth is highly non-linear but implementable.

Last but not least, the risk capacity constraint is significantly different from the leverage constraint. At first glance, it seems to be a particular case of leverage constraint or collateral constraint, $X_t \le f(W_t)$, where $X_t$ represents the dollar invested in the risky asset and $W_t$ the wealth at time $t$. Earlier studies on the leverage constraint include Grossman and Vila (1992), Zariphopoulou (1994), Vila and Zariphopoulou (1997).  Studies on portfolio choice and asset pricing under other dynamic constraints on the control variable $c_t$ or the state variable, $W_t$, include Black and Perold (1992), Dybvig (1995),  El Karoui and Jeanblanc-Picque (1998), Elie and Touzi (2008), Dybvig and Liu (2010), Chen and Tian (2016), Ahn, Choi and Lim (2019), and reference therein. However, most studies on the leverage constraint do not study the situation that $f(W_t)$ is independent on $W_t$, and there are subtle differences as shown in this paper. 
For a comparative purpose, we solve the relevant optimal portfolio choice problem in which a dynamic leverage constraint replaces the risk capacity constraint. We demonstrate that the optimal portfolio under a leverage constraint moves precisely in the stock movement direction, which is a severe concern of the living standard risk in a stressed market period. Therefore, the risk capacity constraint is substantially different from classical leverage constraints, and it can be used to reduce the retiree's living standard risk.

The structure of the paper is organized as follows. Section 2 introduces the model and presents a general optimal investment problem with risk capacity constraint in an infinite time horizon. In Section 3, we show that the constrained region is $(W^*, \infty)$ under certain conditions. The explicit solution of the value function and the consumption-saving strategy are presented in Section 4  (Problem (A)) and Section 5 (Problem (B)), respectively. We present the applications to the retiree's asset allocation problem in Section 6. The conclusion is given in Section 7, and technical proofs are given in Appendix A - Appendix B.

\section{An Optimal Portfolio Choice Problem}
This section introduces a risk capacity constraint and then presents an optimal portfolio choice problem under the risk capacity constraint. Finally, we characterize the value function of this optimal portfolio choice problem in terms of the HJB equation.

\subsection{Investment Opportunities}
 There are two assets in a continuous-time economy. Let $(\Omega, ({\cal F}_t), P)$ be a filtered probability space in which the information flow is generated by a standard one-dimension Brownian motion $(Z_t)$. The risk-free asset (``the bond'') grows at a continuously compounded, constant $r$. We treat the risk-free asset as a numeaire, so we assume that $r = 0$. ${\cal F}_{\infty}$ is the $\sigma$-algebra generated by all ${\cal F}_t, \forall t \in [0, \infty)$.
 
 The other asset  (''the stock index") is a risky asset, and its price process $S$ follows
\be
dS_t = \mu S_t dt + \sigma S_t dZ_t
\ee
where $\mu$ and $\sigma$ are the expected return and the volatility of the stock index.   

\subsection{Investor}
The investor's initial wealth is $W_0$ at time $0$. We simply name ``he" for the investor. The investor is risk-averse and his utility function is denoted by a strictly increasing and concave function $u(\cdot): (0, \infty)  \rightarrow R$ and $u(\cdot)$ satisfies the Inada's condition: $\lim_{W \uparrow \infty} u'(W) = 0$, and $\lim_{W \downarrow 0} u'(W) = 0$. 

\subsection{Risk capacity constraint}

Let $X_t$ be the dollar amount invested in the risky asset at time $t$. Consider a pension portfolio and time $t=0$ represents the beginning of the retirement, and $W_0$ is the  wealth at time $t=0$. For a highly risk averse investor (retiree), we introduce the following constraint
\be
\label{eq:X}
 0 \le X_t \le L, \ \ \ \ \ t \geq 0.
\ee
Here $L = l W_0$ for a positive number $l$. It means that the dollar amount in the market is non-negative (no short-selling) and bounded from above by a percentage of the initial wealth. For example, let $l = 30\%, W_0 = 1, 000,000$, then at most \$300,000 is invested in the stock market during the entire time period. Since this constraint highlights the dollar amount, we call it a {\em risk capacity constraint} and $L$ a capacity. Ottaviani and Sorensen (2015) use this exogenous constraint to study price reaction to information with heterogeneous beliefs.

 Among portfolio constraints in numerous optimal portfolio choice literature, the leverage constraint is mostly related to the risk capacity constraint. That is, $X_t \le k (W_t + L)$, where a positive number $k$ denotes the leverage upper bound of the investment. See classical studies in Grossman and Vila (1992), Vila and Zariphopoulou (1997), Zariphopoulou (1994), and a more recent study in Ahn et al. (2019). Our main insight in this paper (shown below) is that the optimal portfolio under risk capacity constraint behaves significantly differently from the leverage constraint, yielding different implications to retirement portfolio management.


\subsection{An optimal portfolio choice problem }
In this subsection,  we present a portfolio choice problem in which 
an investor's preference is on the consumption path and wealth process.
Specifically, the investor's expected utility is given by 
\be
\label{eq:general}
 \ E \left[  \int_{0}^{\infty} e^{-\delta t} \left\{ \alpha u(c_t) +\beta u(W_t)\right\}dt \right].
\ee
with two nonnegative constants $\alpha$ and $\beta$, and $\alpha + \beta > 0$. 

The optimal portfolio choice problem is to find the optimal trading strategy $(X_t)$ and the consumption rule $(c_t)$ in
\be
\label{value:general}
\bar{V}(W_0,L) \equiv \sup_{(c_t, X_t) \in {\cal A}(W_0, L)} \mathbb{E}\left[ \alpha \int_{0}^{\infty} e^{-\delta t}u(c_t)dt+\beta \int_{0}^{\infty}e^{-\delta t}u(W_t)dt  \right],
\ee
where  ${\cal A}(W_0, L)$ be the set of admissible $(c, X)$ such that (1)
 $c_t$ is ${\cal F}_t$-progressively measurable process, $c_t \ge 0\ a.s., \forall t \ge 0$ and $\int_{0}^{t} c_s ds < \infty\ a.s., \forall t \ge 0$; (2)
 $X_t$ is ${\cal F}_t$-progressively measurable process, and square-integral, $\int_{0}^{t} X_s^2 ds < \infty\ a.s. \forall t \ge 0; $ (3)   $0 \le X_t \le L,\forall t \ge 0$; and (4) the wealth budget constraint,
 \beq
 dW_t = X_t(\mu dt + \sigma dZ_t)  - c_tdt, 
 \eeq
 and $W_t \ge 0\ a.s., \forall t \ge 0$. 
 
In stochastic control literature, there are extent studies on the following general  stochastic control problem,
 \beq
 \max_{c, X} \mathbb{E}\left[\int_{0}^{T} f(t,c_t, W_t, X_t) dt\right],
 \eeq
 without constraint.\footnote{For instance, Bismut (1973)  studies the above general stochastic control problem in a general diffusion process framework and shows duality theorems for a general concave function $f(\cdot)$. However, the characterization of the value function often relies on technical assumptions on the model parameters and the control variables. See Fleming and Soner (2006).} While we consider the {\em additive} specification of the expected utility in (\ref{eq:general}), a {\em multiplicative} specification such as $f(t, c_t, W_t, X_t) = \frac{c_t^{a} W_t^{b} }{1-R}$ is considered in Bakshi and Chen (1996) and Smith (2001).
 
 \subsection{The characterization of the value function}
 
We characterize the value function in (\ref{value:general}) by the following proposition.

\begin{Prop}
\label{pr:general characterization}
The value function $\bar{V}(W)$ is the unique viscosity solution in the class of concave functions of the following HJB equation:
\begin{equation}
\label{eq:hjb-general}
\delta \bar{V}(W) = \max_{ 0 \le X \le L} \left[ \mu X \bar{V}'(W) + \frac{1}{2} \sigma^2 X^2 \bar{V}''(W)  \right]  \\
 + \max_{c \ge 0} \{\alpha u(c) - c \bar{V}'(W)\}+\beta u(W),   (W > 0)
\end{equation}
with  $\bar{V}(0) = \frac{\alpha+\beta}{\delta}u(0)$.   
\end{Prop}

\proof  See Appendix A. \hfill$\Box$

The central point in  Proposition \ref{pr:general characterization} is that the value function is {\em uniquely} characterized in the framework of viscosity solution of the HJB equation, regardless of the smooth property (``ex-ante") of the value function of a portfolio choice problem or not. 
If the utility function $u(\cdot)$ satisfies a global Lipschitz condition, and the control variable $X_t$ takes values in $[l, L]$ for $0 < l < L$, Assumptions 1 - 3 in Strulovici and Szydlowski (2015, Theorem 1) are satisfied; hence, the value function is $C^2$ smooth. It remains open whether the value function is $C^2$ smooth under the risk capacity constraint, or the utility function $u(\cdot)$ does not satisfy a global Lipschitz condition. 

From now on, we consider a widely class of utility function in economic and finance but does not satisfy the global Lipschitz condition. That is,  
\beq
u(W) = \frac{W^{1-R}}{1-R}, R> 0, R \ne 1, (\forall W > 0).
\eeq

{\em Assumption A.}
\begin{eqnarray}
	\delta  > \rho \equiv  \frac{(1-R)\kappa}{R},
\end{eqnarray}
where $\kappa = \frac{\mu^2}{2 \sigma^2}$.  Let
\begin{eqnarray*}
	\lambda^{\infty} = \frac{1}{R}\left[\delta - (1-R)\left(\frac{\mu^2}{2 R\sigma^2}\right)\right].
\end{eqnarray*}
To guarantee the existence of the optimal solution in standard Merton's model, we  impose Assumption A from the next section, $\lambda^{\infty} > 0$.

\section{Characterization of the constrained region}

In this section, we {\em assume the $C^2$ smooth of the value function} to characterize the constrained region under certain conditions.

Specifically, if the value function is $C^2$ smooth, by Proposition \ref{pr:general characterization}, the optimal investment strategy is
\be
\label{eq:A-X}
X^* = \min\left\{ -\frac{\mu}{\sigma^2} \frac{\bar{V}'(W)}{\bar{V}''(W)}, L \right\}.
\ee

Following standard convention in Zariphopoulou (1994), we divide the state space $\Omega = [0, \infty)$ into two regions. On one hand, in the region 
\beq
{\cal U} = \left\{ W > 0: -\frac{\mu}{\sigma^2} \frac{\bar{V}'}{\bar{V}''} \le L  \right\},
\eeq
$X^* < L$, we call ${\cal U}$ the unconstrained region following Vila and Zariphopoulou (1997).\footnote{In the region ${\cal U}$, the constrain $X^* \le L$ in the HJB equation (\ref{eq:hjb-A}) becomes irrelevant. Hence, it is often called the region unconstrained. }
On the other hand, in the region 
\beq
{\cal B} = \left\{ W > 0:  -\frac{\mu}{\sigma^2} \frac{\bar{V}'}{\bar{V}''} > L  \right\},
\eeq
the risk capacity constraint is binding and then $X^* = L$. We call ${\cal B}$ a constrained region.



\subsection{Portfolio choice without  risk constraint}

As a benchmark, we first solve the optimal portfolio choice problem (\ref{value:general}) without the risk capacity constraint; alternatively, $L  = \infty$.

\begin{Prop}
\label{pr:general without constraint} 
In the absence of the risk capacity constraint, the value function in (\ref{value:general}) is
\beq
\bar{V}(W)={A\over 1-R}W^{1-R}
\eeq
where $A$ is a positive constant which will be specified later. The risky asset investment amount is
\beq
X_t = \frac{\mu}{R \sigma^2}W_t.
\eeq
\begin{enumerate}
	\item [(a)] If $\alpha>0,\beta>0$, then $A$ is the unique positive root of the following equation:
 \beq
({\delta \over 1-R}-{\kappa\over R})A={\alpha}^{1\over R}{R\over 1-R}A^{1-{1\over R}}+{\beta\over 1-R},
\eeq
Moreover, the optimal consumption rate is
\beq
c_t^*=({\alpha \over A})^{{1\over R}}W_t.
\eeq
\item [(b)] If $\alpha>0,\beta=0$, then $A=\alpha(\lambda^{\infty})^{-R}$, the optimal consumption rate is $
c_t^*=\lambda^{\infty}W_t.$
\item [(c)] If $\alpha=0$, $\beta>0$, then $A ={\beta\over \delta-\rho}$, the optimal consumption rate is $
c_t^*=0.$
\end{enumerate}
\end{Prop}

\proof See Appendix A. \hfill$\Box$

According to this Proposition \ref{pr:general without constraint}, without the risk constraint constraint, the optimal strategy is a constant proportion of wealth invested in the risky asset and a constant consumption-wealth ratio. As a result, the wealth process is a geometric Brownian motion. Since $W_t$ has a lognormal distribution, the risk capacity constraint fails with a positive positive probability for any time $t > 0$.

\subsection{The unconstrained and constrained region}

Assuming the value function $\bar{V}(W)$ is $C^2$ smooth, we characterize the unconstrained region explicitly under certain conditions in the following result.

\begin{Prop}
	\label{pr:region}
	Assume the value function $\bar{V}(W)$ is $C^2$ smooth, and the following three conditions hold. 
	\begin{enumerate}
	\item [(1)] (``Nontrival unconstrained region") There exists a positive number $W_0 > 0$ such that $(0, W_0) \subseteq {\cal U}$.
\item [(2)] (``Order of value function") There exists two positive numbers $C_0, C_1$ such that $C_0 W^{-R} \le V'(W) \le C_1 W^{-R}$ for all $W \in (0, \infty)$.
\item [(3)] (``Single crossing") The function $g(W) \equiv - \beta W^{-R-1} ( \mu W - \sigma^2 L R)- \alpha^{1/R} {\mu^2\over \sigma^2LR}(\bar{V}(W)')^{1-{1\over R}}$ changes the sign at most one time in the region $(0, \infty)$,
	\end{enumerate}
Then there exists a positive number $W^*$ such that ${\cal U} = (0, W^*]$, and ${\cal B} = (W^*, \infty)$.
\end{Prop} 

\proof See Appendix A. \hfill$\Box$

Proposition \ref{pr:region} is crucial to derive an explicit solution of the general portfolio choice problem (\ref{value:general}). It states that both the unconstrained and constrained region are simply determined by a finite positive number $W^*$ under certain conditions about the value function. Assuming the value function is solved, the optimal investment strategy is $X^* = -\frac{\mu}{\sigma^2} \frac{\bar{V}'(W)}{\bar{V}''(W)}$ if $W \le W^*$, and otherwise, $X^* = L$. Furthermore, if $\alpha \ne 0$, then the optimal consumption rate $c^* = \alpha^{1/R} (\bar{V}'(W))^{-1/R}$. As will be shown below, the number $W^*$ is also essential to derive the value function explicitly.

In Proposition \ref{pr:region}, the ``nontrivial unconstrained region" condition (1) states that the risk constraint condition is satisfied when the wealth is sufficiently small. Its intuition is simple. If the wealth is reasonably small, the investor's investment dollar amount in the risky asset is small as well, then $X_t < L$. The ``order of value function" condition (2) follows from the assumption of the CRRA utility function with order $1-R$. Both condition (1) and (2) are straightforward (See their proofs in some important cases in  Appendix A). Nevertheless, the ``single crossing" condition (3) is more dedicated and essential in characterizing the unconstrained region precisely.

For general values of $\alpha$ and $\beta$, it is hard to check the single crossing condition due to two components in $g(W)$. To see it, we notice that its second component  is strictly monotonic, as the value function increases and concave. Precisely, $(\bar{V}(W)')^{1-{1\over R}}$ is increasing if $R < 1$ and decreasing otherwise. In contrast, its first component, $W^{-R-1} ( \mu W - \sigma^2 L R)$, increases over the region $W \le \frac{\sigma^2 L}{\mu}(R+1)$ and decreases in other region. In total, the function $g(W)$ displays a complicated shape, and the single crossing condition itself depends on some properties of the value function, which is to be determined.

There are two special cases though, $\alpha = 0$ or $\beta = 0$, in which the single crossing condition is satisfied naturally. First, $\alpha = 0$, then $g(W) = - \beta W^{-R-1} ( \mu W - \sigma^2 L R)$ only changes the sign at $W = \frac{\sigma^2 LR}{\mu}$. Second, $\beta = 0$, then $g(W) = - \alpha^{1/R} {\mu^2\over \sigma^2LR}(\bar{V}')^{1-{1\over R}} < 0$.

Our objective is to explicitly investigate the risk constraint's implications by giving an analytical expression of the optimal consumption-investment policy. Therefore, we next  focus on these two situations, $\alpha = 0$ or $\beta = 0$. Specifically, we explicitly solve two portfolio choice problems. The first one is given by
\be
\text{Problem (A): }
U(W_0,L) \equiv \sup_{(c_t, X_t) \in {\cal A}(W_0, L)} \mathbb{E}\left[ \int_{0}^{\infty} e^{-\delta t}u(c_t)dt  \right],
\ee
in which the value function is written by $U(W_0, L)$ to highlight the risk constraint level $L$. For any $0 < L_1 < L_2$, it is evident that $U(W, 0)  \le U(W, L_1) \le U(W, L_2) \le U(W, \infty)$, where $U(W, \infty)$ is the value function in Merton's model without the constraint on $(X_t)$. If there is no confusion, we will write $U(W)$ to represent $U(W,L)$ in this paper. When $X_t \le f(W_t)$, in particular, $X_t \le k (W_t + L),$ Problem (A) is studied in Vila and Zariphopoulou (1997), Zariphopoulou (1994). 

The second one is given by 
\be
\label{eq:problem}
\text{Problem (B): } 
V(W_0,L) = \max_{(X)} \mathbb{E}\left[ \int_{0}^{\infty} e^{- \delta t} u(W_t) dt \right].
\ee
Here, we use $V(W,L)$ in Problem (B) to denote the value function to distinguish from $U(\cdot)$. Similarly, we do not distinguish $V(W,L)$ with $V(W)$ if it is evident.

In the following sections,  we derive the explicit solution of Problem (A) and Problem (B), respectively.

\section{Solution to Problem (A)}


For Problem (A), the smooth property of the value function follows from a difficult theorem of Zariphopoulou (1994).\footnote{Precisely, Zariphopoulou (1994) investigates the constraint that $X_t \le f(W_t)$ almost surely, where $f(x): [0, \infty)  \rightarrow [0, \infty)$ is increasing, concave and satisfies the global Lipschitz condition. Her dedicate argument goes through if $f(\cdot)$ is a positive constant.} Zariphopoulou (1994, Theorem 1.1) shows that 
 $U(W)$ is the unique $C^2$ smooth function of the following HJB equation:
\begin{equation}
\label{eq:hjb-A}
\delta U(W) = \max_{ 0 \le X \le L} \left[ \mu X U'(W) + \frac{1}{2} \sigma^2 X^2 U''(W)  \right]  \\
 + \max_{c \ge 0} \{u(c) - c U'(W)\},  (W > 0)
\end{equation}
with  $U(0) = \frac{u(0)}{\delta}$.   

Since the value function $U(W)$ is $C^2$ smooth, both the unconstrained and constrained region are well-defined. On one hand, in the unconstrained region, the HJB equation (\ref{eq:hjb-A}) reduces to the following ordinary differential equation
\be
\label{eq:hjb1}
\delta U = \frac{R}{1-R}(U')^{-(1-R)/R} - \kappa \frac{(U')^2}{U''}.
\ee
On the other hand, in the constrained region ${\cal B}$, the HJB equation reduces to  another ordinary differential equation
\be
\label{eq:hjb2}
\delta U = \frac{R}{1-R}(U')^{-(1-R)/R} + \mu L U' + \frac{1}{2} \sigma^2L^2 U''.
\ee

\begin{Prop}
\label{pr:characterization2}
There is a positive number $W^*$ such that ${\cal U} = (0, W^*]$ and ${\cal B} = (W^*, \infty)$.
\end{Prop}

\proof See Appendix A. \hfill$\Box$

Proposition \ref{pr:characterization2} is crucial to derive an explicit solution to Problem (A). It not only characterizes the constrained region but also reduces Problem (A) to determine the threshold number $W^*$ next.


Define
\begin{eqnarray*}
\omega^{+} = {R\over\kappa}(\kappa+\delta),
\end{eqnarray*}
then, for any $R >0, R \ne 1$, Assumption A implies that  $
\omega^{+} > \max(R, 1).$

Given a real number $m$, define two auxiliary function $H(C)$ and $J(C)$ as follows.
\be
\label{eq:H}
H(C) = \frac{1}{\lambda^{\infty}}\left(C + m C^{\omega^{+}}\right),
\ee
\be
J(C) = \frac{1}{\lambda^{\infty}} \left\{ \frac{C^{1-R}}{1-R} + m\frac{\omega^{+}}{\omega^{+} - R} C^{\omega^{+} - R}   \right\},
\ee

\begin{Prop}
\label{pr:solution}
 There exists unique real numbers $\{C^*, m\}$ which satisfies the following two equations 
\be
\label{eq:first-order}
 C^* H'(C^*) = \frac{\sigma^2 RL}{\mu},
\ee
and
\be
\label{eq:value}
J(C^*) = K(W^*),
\ee
where $K(x)$ satisfies the following second-order ordinary differential equation
\begin{eqnarray}
\label{eq:K}
\frac{1}{2} \sigma^2 L^2 K''(x) = \delta K(x) - \frac{R}{1-R} (K'(x))^{-(1-R)/R} - \mu L K'(x),
\end{eqnarray}
with $K'(W^*) = (C^*)^{-R}$, and $K(W)$ has the same order as $W^{1-R}$ when $W \rightarrow \infty$.
 Moreover, $W^* = H(C^*)$.
\end{Prop}

\proof See Appendix A. \hfill$\Box$

Proposition \ref{pr:solution} is understood as follows. In the unconstrained region $(0, W^*]$, the relationship between wealth and optimal consumption rate is characterized by $W = H(C)$, a strictly increasing function.\footnote{It is a well known fact the function $H(\cdot)$ is strictly increasing, since $U(\cdot)$ is concave and $U_{WW} = -R C^{-R-1} \frac{\partial C}{\partial W} = -R C^{-R-1}/H'(C)$. Furthermore, if the optimal consumption rate is a concave function of the wealth (Carroll and Kimball (1996)), then $\frac{\partial^2 C}{\partial W^2} = - \frac{H''(C)}{H'(C)^2} \frac{\partial C}{\partial W} < 0$ implies that $H(\cdot)$ is a convex function.}
The value function $U(W)$ is determined by $J(C)$. Therefore, the value function is characterized with the auxiliary parameter $C$. Moreover, the optimal investment policy is 
\be
X^*(W) = \frac{\mu}{R \sigma^2} C H'(C).
\ee
Since $C H'(C)$ is increasing with respect to $C$, $X^*(W)$ is a well-defined increasing function of the wealth in the unconstrained region. Therefore, $X^*(W) < L$ holds for any wealth $W < W^*$. 

In the constrained region $(W^*, \infty)$, the optimal investment policy is straightforward,  $X^* = L$, the investor places $L$ dollars in the risky asset as long as the portfolio value is greater than $W^*$. The value function $U(W) = K(W)$ satisfies the following ordinary differential equation
\begin{eqnarray}
\label{eq:K}
\frac{1}{2} \sigma^2 L^2 K''(x) = \delta K(x) - \frac{R}{1-R} (K'(x))^{-(1-R)/R} - \mu L K'(x).
\end{eqnarray}
with the boundary conditions $K(W^*)= J(C^*)$ and $K'(W^*)= (C^*)^{-R}$. These two boundary conditions of the function $K(x)$ are exactly the value-matching and smooth-fit condition at $W^*$. However, these two conditions alone are not sufficient yet to characterize uniquely the function $K(x)$ as $W^*$ is to be determined. Therefore, we need a boundary condition at $W \rightarrow \infty$, that is, the order of $K(x)$, according to standard theory of ordinary differential equation. 
Finally,  the optimal consumption rate is $c^* = K'(W)^{-\frac{1}{R}}$. 

\section{Solution to Problem (B)}

In contrast to Problem (A), the value function of Problem (B) is not known to be $C^2$ smooth under the risk constraint. Therefore, in solving Problem (B), we need to simultaneously investigate the smooth property of the value function. We follow three steps. First, we explicitly characterize the constrained and unconstrained region of the risk capacity constraint, assuming the value function is $C^2$ smooth. This characterization of the constrained region follows from Proposition \ref{prop:region}. Second, we provide an explicit expression of the value function, still assuming the $C^2$ smooth property of the value function. Third, we present the sufficient and necessary condition of the $C^2$ smoothness of the value function, given the expression of the value function in the second step. 


{\em Assuming the value function $V(W)$ is $C^2$ smooth,}
the value function on the unconstrained region satisfies 
\be
\label{eq:U}
 \delta V(W) = u(W) + \kappa \frac{(V'(W))^2}{-V''(W)}.
\ee

Similarly, on the constrained region, $V(\cdot)$ satisfies a second-order linear ODE
\be
\label{eq:BB}
\delta V(W) = u(W)  + \mu L V'(W) + \frac{1}{2} \sigma^2 L^2 V''(W).
\ee

In the above definition of the region ${\cal U}$ and ${\cal B}$, we assume that $C^2$ smoothness of the value function ex ante.  Without knowing the $C^2$ smooth property of the value function, we can still directly investigate  the ordinary differential equation (\ref{eq:U})  - (\ref{eq:BB}). Later, we study these two ordinary differential equations and verify the $C^2$ smooth property of the value function under certain conditions.


\begin{Prop}
\label{prop:region}
Assume $V(W)$ is $C^2$ smooth, then there exists a positive number $W^*$ such that ${\cal U} = (0, W^*]$ and ${\cal B} = (W^*, \infty)$.
\end{Prop}

\proof See Appendix A. \hfill$\Box$

By Proposition \ref{pr:general without constraint}, the number $W^*$ is a finite number. 
Similar to Proposition \ref{pr:characterization2} for Problem (A), the characterization of ${\cal U}$ and ${\cal B}$ in Proposition \ref{prop:region} reduces the solution to a number $W^*$.


We next derive the sufficient and necessary condition of the $C^2$ smoothness of the value function and simultaneously characterize the threshold $W^*$. 
Define two real numbers 
\be
\beta_1 = \frac{- \mu + \sqrt{\mu^2 + 2   \delta \sigma^2}}{\sigma^2 L},
\beta_2 = \frac{- \mu - \sqrt{\mu^2 + 2  \delta \sigma^2}}{\sigma^2 L},
\ee
where $\beta_1$ and $\beta_2$ are two roots of the following quadratic equation 
\beq
\frac{1}{2}\sigma^2 L^2 \beta^2 + L \mu \beta  - \delta = 0,
\eeq
and $\beta_1 > 0 > \beta_2$.


Let
\begin{eqnarray*}
V_{0}(W) & = & \frac{2}{(\beta_1 - \beta_2)(1-R)\sigma^2 L^2} \\
&& \times \left\{ e^{\beta_2 W} \int_{0}^{W} x^{1-R} e^{-\beta_2 x}dx -
e^{\beta_1 W } \int_{0}^{W} x^{1-R} e^{-\beta_1 x}dx \right\}.
\end{eqnarray*}
The function $V_0(W)$ is a well-defined smooth function for $W > 0$. We recall the expression of lower incomplete Gamma function,
\beq
\gamma(s,x)=\int_0^{x} t^{s-1}e^{-t}dt, Re(s) > 0,
\eeq
which is well-defined for all real number $x>0$. Therefore, $V_0(W)$ is well-defined for any $0 < R < 2$.\footnote{We refer to Appendix B for basic properties of the incomplete Gamma function.}

Given $V_0(\cdot)$, the following  two functions $C(W)$ and $D(W)$ are well-defined.
\be
\label{eq:C*}
C(W) ={-{2\over(\beta_1-\beta_2)(1-R)\sigma^2L^2}\beta_1^{R-1}\Gamma(2-R)e^{\beta_1 W}[\mu+L\sigma^2\beta_1]-\mu V_0'(W)-\sigma^2 LV_0''(W)\over \sigma^2 L(\beta_2)^2e^{\beta_2 W}+\mu \beta_2 e^{\beta_2 W}},
\ee
and
\be
\label{eq:g*}
D(W) =\left\{ \frac{2}{(\beta_1-\beta_2)(1-R)\sigma^2L^2}(\beta_1)^{R-1}\Gamma(2-R)e^{\beta_1 W}+C(W) \beta_2e^{\beta_2 W}+V_0'(W) \right\}^{-\frac{1}{R}}.
\ee
Given any $W > 0$, we define a function $G(g), 0 \le g \le D(W),$ by the following second-order ordinary differential equation,
\be
\label{eq:G}
G''(g)={R\over \kappa}g^{-1}G'(g)\left[ \delta -\rho  -g^{R}(G(g))^{-R}\right]
\ee
with boundary condition $G(0) = 0, G(D(W)) = W$ and $G'(D(W))={LR\sigma^2\over \mu}(D(W))^{-1}$ (See King, Billingham, and Otto (2003) for the properties of second-order ordinary differential equation). 

Finally, we define one equation of the variable $W$ as follows.
\be
\label{eq:W*}
\frac{u(0)}{ \delta }+\int_0^{D(W)} g^{-R}G'(g)dg={2\over (\beta_1-\beta_2)(1-R)\sigma^2L^2}(\beta_1)^{R-2}\Gamma(2-R)e^{\beta_1 W}+C(W) e^{\beta_2 W}+V_0(W).
\ee
In what follows, we are interested in one particular function $G(\cdot)$ that is defined by one specific number $W^*$.

\begin{Prop}
\label{prop:solution} Assume $0 < R < 2$.

(``{\em Necessary}'')  If $V(W,L)$ is $C^2$, then there exists a unique positive solution $W^*$ of Equation (\ref{eq:W*}), and the value function is given by
\begin{eqnarray} 
\label{eq:solution}
\tilde{V}(W,L)=
\begin{cases}
\frac{u(0)}{\delta}+\int_0^{G^{-1}(W)}g^{-R}G'(g)dg, & W\leq W^*\cr
\frac{2}{ (\beta_1-\beta_2)(1-R)\sigma^2L^2}(\beta_1)^{R-2}\Gamma(2-R)e^{\beta_1 W}+C(W^*) e^{\beta_2 W} + V_{0}(W), &W>W^*. \end{cases}
\end{eqnarray}
Here, the function $G(g)$ is defined by the real number $W^*$. 

(``{\em Verification}'') Assume the existence of a positive solution $W^*$ of Equation (\ref{eq:W*}). Moreover, in the region $0 \le g \le g^* = D(W^*)$,  $G(g)$ is increasing, 
\beq
G(g) \ge \left( \delta - \rho + \frac{\kappa}{R} \right)^{-\frac{1}{R}}g,
\eeq
 and  $ \mu \tilde{V}_{W}(W,L) + L^2 \tilde{V}_{WW}(W,L) \ge 0$ for all $W \ge W^*$. Then the number $W^*$, as a positive solution of Equation  (\ref{eq:W*}) is unique, the value function $V(W,L)$ is $C^2$ smooth and $V(W,L) = \tilde{V}(W, L)$ in (\ref{eq:solution}).

\end{Prop}

\proof See Appendix A. \hfill$\Box$

Proposition \ref{prop:solution}  presents a sufficient and necessary condition of the $C^2$ smoothness of the value function and solves Problem (B) explicitly. It also presents a closed-form expression of the value function and the optimal strategy in terms of $W^*$ and the auxiliary function $G(\cdot)$.  By its construction, if there is one positive solution of Equation (\ref{eq:W*}), then the solution $W^*$ is unique, and the unconstrained region and the constrained region are separated by $W^*$. Moreover, the value function $V(W,L)$ is a $C^2$ smooth function of the HJB equation and given by the expression (\ref{eq:solution}) in Proposition \ref{prop:solution}. Conversely, if $V(W,L)$ is $C^2$ smooth, then there exists a unique solution of Equation (\ref{eq:W*}). The $C^2$ smoothness of the value function is essentially the existence of a positive solution of the nonlinear equation (\ref{eq:W*}) of one variable.

If $L=\infty$, it reduces ${\cal U} = (0, \infty)$ and an empty region ${\cal B}$. By Proposition \ref{pr:general without constraint}, the function $G(\cdot)$ is a linear function.  
For any $L < \infty$, the function $G(\cdot)$ is highly non-linear, and its non-linearity is equivalent to the non-myopic property of the optimal strategy, as will be explained in the next section. 

\section{Applications to optimal retirement portfolio}
In this section, we present applications to the optimal portfolio for retirees. Our objective is to explain the implications of the risk capacity constraint and demonstrate its substantial difference from the leverage constraint.

There are several distinct features in the retiree's portfolio choice problem compared with a standard investor before retirement. (1) The retiree has a fixed cash flow from his social security account post-retirement.\footnote{
See www.ssa.gov for the social security system in the U.S.A. There are similar social security systems in Europe and Canada.}
 (2) He has no labor income. (3) He has a mortality risk, and (4) he becomes more risk-averse than before retirement because he has concerns about the market downturn and has no sufficient time to wait for the market return, leading to a substantial decline in his living standards. We formally incorporate these features into the optimal portfolio choice problems as follows.

The retirement date is zero. The retiree's initial wealth is $W_0$ at the retirement date.  Since the retiree faces his mortality risk, the investment time-horizon is uncertain, neither a fixed finite time nor infinity. We assume that the investor's death time $\tau$ has an exponential distribution with mean $\lambda$, that is, the probability of the retiree surviving in the next $t$ years is $e^{- \lambda t}$. The investor's average lifetime is $\frac{1}{\lambda}$, and the variance of his lifetime is $\frac{1}{\lambda^2}$. For example, if $\lambda = 0.05$, an ordinary retiree who retires at 65 is likely to die at 85 years old. $\tau$ is independent of the information set ${\cal F}_{\infty}$.

We first consider an optimal portfolio choice problem with random maturity (see Blanchet-Scaillet, et al. 2008; Chen, Fu, and Zhou, 2020) as follows,
\be
\label{eq:problem1}
\max_{X, c} \mathbb{E}\left[\int_{0}^{\tau} e^{-\delta t} u(c_t)dt \right].
\ee
Here the wealth process $W_t$ satisfies $dW_t = X_t(\mu dt + \sigma dZ_t)  - c_tdt, W_0 = W$, $W_t \ge 0\ a.s., \forall t \ge 0$, and the risk capacity constraint $0 \le X_t \le L, \forall t \le \tau$.\footnote{When the retiree receives a constant social security stream, the wealth equation becomes $dW_t = X_t(\mu dt + \sigma dZ_t)  - (c_t - y_0)dt$, $y_0$ represents the social security with continuously compounding, the discussion in Problem (A) can be applied if we consider $u(c_t - y_0)$ in Problem (A). The reason is as follows. If the social security is sufficient, there is no need to withdraw from the pension portfolio. Therefore, the retiree focuses on the difference $c_t - y_0$ in the withdraw decision. To sharpen our intuition of the risk capacity constraint, we omit the social security or other fixed cash-flow income. We also ignore medical costs or other costs in the portfolio choice decision.}
 In this problem, the investor finds the best withdraw (consumption) rate whereas the inheritance wealth is not a concern. Because the stopping time $\tau$ is independent from the equity market information,
\begin{eqnarray*}
\mathbb{E}\left[\int_{0}^{\tau} e^{-\delta t} u(c_t)dt \right] &=& \mathbb{E}\left[\int_{0}^{\infty} e^{-\delta t}1_{\tau \ge t} u(c_t)dt \right]\\
& = & \mathbb{E}\left[\int_{0}^{\infty} \mathbb{E}[e^{-\delta t}1_{\tau \ge t} u(c_t)|{\cal F}_{\infty}]dt \right]\\
&=& \mathbb{E}\left[\int_{0}^{\infty} e^{-\delta t} u(c_t) e^{-\lambda t}dt \right]
\end{eqnarray*}
where we make use of the fact that $P(\tau \ge t) = e^{-\lambda t}$. Therefore, this optimal retirement portfolio reduces to Problem (A) in which the the subjective discount factor is replaced by $\delta + \lambda$. It is thus natural to assume that $\delta + \lambda < 1$ and then $\delta + \lambda > \rho$.

Alternatively, the second  optimal retirement portfolio problem for the retiree at time zero is as follows (see Liu and Loewenstein, 2002),
\be
\label{eq:problem2}
J(W, L) = \max_{(X)} \mathbb{E}\left[  e^{-\delta \tau} u((1-\alpha)W_{\tau}) \right]
\ee
where $\delta$ is the retiree's subjective discount factor, $\alpha$ is the inheritance tax rate  of the wealth. 
 $X_t$ satisfies the risk capacity constraint. 
%
Given the distribution of $\tau$, and the independent assumption between $\tau$ and ${\cal F}_{\infty}$, using the same derivation as in the first problem, 
 the retiree's optimal retirement portfolio problem (\ref{eq:problem2}) is reduced to
\be
\label{problem:merton}
J(W, L) = \max_{(X)} \mathbb{E}\left[ \int_{0}^{\infty} e^{-( \lambda + \delta) t}   u((1- \alpha) W_t) dt \right].
\ee
 Assuming $u(W) = \frac{W^{1-R}}{1-R}, R >0, R \ne 1$, and using its scaling property,  
 $$ u((1-\alpha) W_t) =  (1-\alpha)^{1-R} u(W_t), $$
 we have 
 \be
J(W, L)  = (1-\alpha)^{1-R} \max_{(X)} \mathbb{E}\left[ \int_{0}^{\infty} e^{-(\lambda + \delta) t} u(W_t) dt \right].
\ee

A general expression of the retirement portfolio problem elaborates these two cases. If both the consumption/withdraw and the inheritance wealth are considered together, \footnote{In practice, a constant consumption rate is often fixed. For instance, a standard withdrawal rate is between 4\% to 5\%. See Bengen (1994). Then the problem reduces to Problem (B).}
\beq
\mathbb{E}\left[  \int_{0}^{\tau} e^{-\delta t} u(c_t)dt + e^{-\delta \tau} u((1-\alpha)W_{\tau}) \right] = 
\mathbb{E}\left[ \int_{0}^{\infty} e^{-(\lambda + \delta) t} \left\{ u(c_t) + (1-\alpha)^{1-R} u(W_t) \right\}dt \right].
\eeq
It reduces to the optimization problem (\ref{value:general}) studied in Section 2 and Section 3.

\subsection{Alternative portfolio choice under a leverage constraint}

We first present the solution to a relevant portfolio choice problem by replacing the risk capacity constraint with a leverage constraint for a comparison purpose. Specifically, we assume $X_t \le b W_t, \forall t$. 
We use a predetermined number of $b$  to represent the highest possible percentage of wealth invested in the risky asset. For instance, $b = 0.7$ means at most 70 percent of the portfolio is invested in the risky asset.  Define
\be
V^{b}(W) = \sup_{(X)} \mathbb{E}\left[ \alpha \int_{0}^{\infty} e^{-\delta t}u(c_t)dt+\beta \int_{0}^{\infty}e^{-\delta t}u(W_t)dt  \right],
\ee
where the risk capacity constraint is replaced by $X_t \le b W_t, \forall t$. By Proposition \ref{pr:general without constraint}, we assume that  $b < \frac{\mu}{R \sigma^2}$. Otherwise, $V^{b}(W)$ is solved by Proposition \ref{pr:general without constraint} for all $b \ge \frac{\mu}{R \sigma^2}$. 

\begin{Prop}
	\label{prop:alternative}
	Under the constraint that $X_t \le b W_t$ and $b < \frac{\mu}{R \sigma^2}$, 
	then
	\be
	V^{b}(W) = B u(W),
	\ee
	 the optimal strategy is $X_t = b W_t$, and the optimal consumption rate is $c_t = \left(\frac{\alpha}{B}\right)^{1/R}W_t$. Here, $B$ is a unique positive number satisfying
	\beq
	\left[\delta + (1-R)(\frac{1}{2}\sigma^2 b^2 R - \mu b)\right] B = \beta + \alpha^{1/R} R B^{1-\frac{1}{R}}.
	\eeq	
\end{Prop}

\proof See Appendix A. \hfill$\Box$

Proposition \ref{prop:alternative} states that a constant percentage strategy $X_t = b W_t$ is an optimal policy under a leverage constraint. 
For $\alpha = 1$ and $\beta = 0$,  Proposition \ref{prop:alternative} reduces to Vila and Zariphopoulou (1997, Proposition 4.2). For $\alpha = 0, \beta = 1$, $B = {1\over \delta + (1-R)({1\over 2}\sigma^2b^2R-\mu b) }$. Under the leverage constraint $X_t \le b W_t$, the general portfolio choice problem in (\ref{value:general}) has a similar optimal strategy and a constant consumption-wealth ratio (for $\alpha \ne 0)$. The wealth process is a geometric Brownian motion. As a consequence, the portfolio wealth at any time has a lognormal distribution.

\subsection{Optimal strategy}

We start with the optimal investing strategy in Problem (A) - (B).
Its explicit expression is given by the next result.

\begin{Cor}
\label{prop:strategy}
The optimal investment strategy in Problem (B) is
\begin{eqnarray} 
X(W)=
\begin{cases}
\frac{\mu}{R\sigma^2}g G'(g), & W\leq W^*,\cr
L, &W>W^*. \end{cases}
\end{eqnarray}
Th expression of the optimal investment strategy for Problem (A) is the same if $g G'(g)$ is replaced by $CH'(C)$.
The optimal portfolio strategies in both Problem (A) and Problem (B)  are not the myopic. 
\end{Cor}

Since the auxiliary parameter, $C$ in Proposition \ref{pr:solution} and $g$ in Proposition \ref{prop:solution}, represents the optimal consumption rate $c^*$ in the unconstrained region $\{W \le W^* \}$, the optimal investment strategy shares the same expression in Corollary \ref{prop:strategy}. If the optimal strategy were  the myopic strategy in the sense that $X_{t}  = \min\left\{ \frac{\mu}{R \sigma^2} W_t, L \right\},$ then the function $G(\cdot)$ or $H(\cdot)$ would be a linear function, and $W^* = \frac{L R \sigma^2}{\mu}$. For Problem (B), it is impossible by the definition of the function $G(\cdot)$, and Equation (\ref{eq:W*}) fails for $W^* = \frac{L R \sigma^2}{\mu}$ because the left side is a polynomial function while the right hand is virtually an incomplete Gamma function.  Intuitively, the risk capacity constraint in the future affects the investment decision even though the constraint is not binding instantly. By the same reason, in Problem (A), the parameter $m \ne 0$ in function $H(\cdot)$ since the future risk capacity constraint affects the decision even for $W < W^*$. In other words, while the risk capacity constraint is not binding, the investment strategy is affected by the fact that the constraint may be binding in the future. The major difference between Problem (A) and Problem (B) is the characterization of the threshold $W^*$ in terms of different non-linear equation. 

%

As a numerical illustration, we plot the auxiliary function $G(\cdot)$ and the investing strategy $X(W)$ in Problem (B). We choose the risk premium $\mu = 0.10$ to consistent with the market data of S \& P 500 between 1948 and 2018. We choose $\lambda= 0.07$ to consistent to approximately 15 years of life after retirement. Assuming the initial retirement portfolio worth 1 million, we choose $700,000$ as the maximum dollar amount in the stock market. 
Let $\sigma = 30\%$. The number $\sigma$ is slightly higher than the calibration of the market index since our purpose is to highlight the high likelihood of the market downturn, which is a big concern for the retiree. We choose $R = 0.5$. By calculation, the expected value for retirement level is $W^* = 490,235$ in Equation (\ref{eq:W*}).   

As shown in Figure 1 and Figure 2, since $G(\cdot)$ is {\em not} a linear function, the strategy is not a myopic one. By the same reason, $X(W)$ as a function of  the wealth is {\em not} $C^1$ since
\beq
\frac{\partial X(W)}{\partial W}|_{W = W^*-} = \frac{\mu}{R \sigma^2} \frac{(g G'(g))'}{G'(g)} = \frac{\mu}{R \sigma^2} \left( 1 + \frac{g G''(g)}{G'(g)}\right) \ne 0.
\eeq
The percentage of wealth in the risky asset, $\frac{X(W)}{W}$, can be analyzed similarly. In the constrained region, $W \ge W^*$, the percentage of wealth is $\frac{L}{W}$. The larger the wealth, the smaller percentage of wealth is invested in the stock market. On the other hand, in the  unconstrained region, $\frac{X(W)}{W} = \frac{\mu}{R \sigma^2} \frac{g G'(g)}{G(g)} $. This function also decreases with respect to the wealth as shown in Figure 3. 

A decreasing percentage of wealth invested in stocks is different from the surveys of the household before retirement. Wachter and Yogo (2010) explain the increasing portfolio shares in the wealth. In contrast, we show the decreasing effect of the risk capacity constraint on the portfolio share $\frac{X(W)}{W}$. In the retirement period, the more wealth, the less portfolio share if the retirees are concerned about the living standard risk.\footnote{However, we admit that our model mainly applies to a median household, not for the wealthy. Even though the wealthiest household still saves more in the portfolio, they are heavily skewed toward risky assets such as their own privately-held business and different preference with the median household. See Carroll (2002).} Both Figure 2 and Figure 3 show that the optimal portfolio strategy displays a strong risk-averse feature by comparing with the benchmark model without the risk capacity constraint. Figure 4 displays a similar comparison when the risk aversion parameter $R = 1.5$.

Similarly, we illustrate the the effect of the risk capacity constraint and its comparison with the leverage constrain for Problem (A) in Figure 5. The Benchmark represents Merton's classical model in which a constant proportion invested in the equity market. ``VZ" denotes the optimal portfolio strategy solved in Vila and Zariphopoulou (1997)
under a leverage constraint $X_t \le \frac{1}{2} \frac{\mu}{R \sigma^2}W_t$. At first glance,  It seems that the optimal investment stratery $X$ is $C^1$ at $W^*$, which is mainly because the function $H$ (see (\ref{eq:H})) in Problem (A) has better smoothness than $G$ (see (\ref{eq:G})) in Problem (B).  However,  we point out that this is not true.  Indeed,  by Lemma C.4 in Vila and Zariphopoulou (1997),  in the unconstrained region ${\cal U}$, we have
\beq
{dX\over dW}={\mu \over R\sigma^2}{C+m(\omega^+)^2C^{\omega^+}\over C+m\omega^+ C^{{\omega}^+}}
\eeq
If $X$ is $C^1$ at $W^*$, then we must have
\beq
C^*+m(\omega^+)^2(C^*)^{\omega^+}=0
\eeq
which contradicts what we derived in Proposition \ref{pr:solution}.

\subsection{Wealth  process}

Given the optimal strategy characterized in Corollary \ref{prop:strategy}, the optimal wealth process in Problem (B) is uniquely determined by (for $W = G(g)$)
\be
\label{eq:sdeW}
dW_t = \min\left\{ \frac{\mu}{R \sigma^2} g G'(g), L \right\} (\mu dt + \sigma dZ_t), W_0 = W > 0.
\ee
 It can be shown that the stochastic differential equation (\ref{eq:sdeW}) has a strong solution. Therefore, we can  directly analyze the portfolio by the stochastic differential equation (SDE) (\ref{eq:sdeW}). We obtain a similar SDE of the wealth process in Problem (A).
 
 The portfolio dynamic is as follows.  Assuming wealth $W_t = W^*$ at a time $t$ from below, then in the instant time period, $[t, t+\delta t]$,
$W_{t, t + \delta t} = W_{t} + L (\mu \delta t + \sigma \sqrt{\delta t} \zeta), $
and $
S_{t+\delta t} = S_{t} + S_t \left( \mu \delta t + \sqrt{\delta t} \zeta \right),$
where $\zeta$ is a standard normal variable.  In a good scenario of the stock market, $S_{t + \delta t} \ge S_{t}$, that is,  $\mu \delta t + \sigma \sqrt{\delta t} \zeta > 0$, then $W_{t+\delta t} \ge W_{t}$, so 
the same dollar amount $L$ is still invested in the stock market. If the market drops in the period $[t, t + \delta t]$, $S_{t+\delta t} < S_{t}$, then $W_{t+\delta t} < W^*$, the portfolio value reduces and is smaller than the threshold $W^*$, then  a new dollar amount, $\frac{\mu}{-R \sigma^2} \frac{V'(W_{t+\delta t})}{ V''(W_{t + \delta t})}$, is invested in the stock market. This process continuous between the unconstrained region and the constrained region. 

The retirement portfolio's return process is 
\beq
\frac{dW_t}{W_t} = \frac{ \min\left\{ \frac{\mu}{R \sigma^2} g G'(g), L \right\}}{W_t} \left( \mu dt + \sigma dZ_t \right). 
\eeq
Therefore, the instantaneous variance, $Var\left[\frac{dW_t}{W_t}\right]$, converges to zero as $W \rightarrow \infty$. When the wealth is sufficiently high, the risk of the portfolio is very small so the retiree is able to resolve the living standard risk, regardless of possible market downturn. More importantly, the instantaneous covariance between $\frac{dW_t}{W_t}$ and $\frac{dS_t}{S_t}$ is
\be
Cov\left(\frac{dW_t}{W_t}, \frac{dS_t}{S_t} \right)  = \frac{X(W_t)}{W_t} \sigma^2 \rightarrow 0, \text{ as } W_t \rightarrow \infty.
\ee
Hence, the portfolio is virtually independent from the stock market if the portfolio value is large enough. The same result holds for Problem (A) by replacing $g G'(g)$ by $C H'(C)$ and $W = H(C)$.

The next result summarizes our discussion.

\begin{Cor}
Under the risk capacity constraint in Problem (A) and Problem (B), the optimal portfolio is {\em virtually independent} of the stock market if the retirement portfolio value is large enough.
\end{Cor}







\subsection{Implications}

In this section, we explain several implications to the retirement portfolio from our results. 

{\em First}, the retiree needs to invest in the stock market since the all-safe strategy is too conservative to sustain the spending given longevity risk. Indeed, Vanguard (2018), among others, suggests that investing after retirement is both necessary and vital. {\em Second}, we demonstrate that the risk capacity constraint captures the retiree's living standard risk, and the optimal portfolio under the risk capacity constraint is a reasonable retirement strategy. Specifically,  if the retirement portfolio value is not high enough, the retiree should invest some money in the stock market to increase the growth rate. However, when the portfolio value is high enough, the retiree implements a ``contingent constant-dollar amount strategy"  by only placing  $L$ dollar of the portfolio in the stock market as long as the portfolio value is higher than $W^*$. {\em Third}, under the risk capacity constraint, the higher the portfolio value, the smaller the wealth in the stock market. As a result, the portfolio can reduce the living standard risk because its return is asymptotically independent of the stock market for a high level of portfolio value. {\em Fourth}, the risk capacity constraint and the leverage constraint yield different investment strategies. The generating retirement portfolio is perfectly correlated to the stock market by implementing a leverage constraint, so the retiree faces a substantial market risk. {\em Finally}, these features are robust regardless of the retiree focuses on consumption (Problem (A)) or the inheritance wealth to her heirs (Problem (B)). 
In interpreting this optimal strategy and saving policy, the number $W^*$ is crucial. Given its unique feature in the investment strategy, this number measures the expected lump sum of the spending in the retirement period.

It is interesting to see the effect of the capacity $L$ to the threshold $W^*$, and we write it as $W^*(L)$. By its definition, we write
\be
X(W^*(L), L) = L,
\ee
where $X(\cdot, L)$ denotes the dollar amount function of wealth in the unconstrained region. By the chain rule in Calculus, we obtain
\be
\frac{\partial X}{\partial W} \frac{\partial W^*(L)}{\partial L} + \frac{\partial X}{\partial L} = 1.
\ee
Therefore, $\frac{\partial W^*(L)}{\partial L}  > 0$ if (a) $\frac{\partial X}{\partial W}  > 0$, and (b) $\frac{\partial X}{\partial L} < 1$. Here, the condition (a) means the monotonic property of the investment in the risky asset in the unconstrained region.\footnote{In Problem (A), if the function $CH'(C)$ is increasing, then $X^*(W)$ is increasing to the wealth $W$. Similarly, if the function $g G'(g)$ increases, we obtain the monotonic property (a) in Problem B. Both follow from the concavity of the consumption rate in Carroll and Kimball (1996). The result still holds if the decreasing rate of $H'(\cdot)$ or $G'(\cdot)$ is bounded from above by the constant 1, even though the concavity of the consumption rate fails.} The condition (b) states that the marginal effect of the capacity to the dollar investment is less than one. While it seems difficult to prove condition (a) and condition (b) rigorously,  their intuitions are appealing. 
For instance, the condition (b) roughly means that
\beq
X(W, L+\epsilon) < X(W, L) + \epsilon.
\eeq
To see it, let $X(W,L)$ be the optimal investment for the capacity $L$, and we now increase the capacity by $\epsilon$. Since the capacity is the maximum possible dollar amount invested in the market, the dollar invested in the equity with the new capacity level $L + \epsilon$ should be bounded from above by the sum of $X(W,L)$ and $\epsilon$. Figure 6 also numerically demonstrates this property of $W^*(L)$ to the capacity $L$.


Choosing the parameter $L$ or $l = L/W_0$ is practically interesting to implement the risk capacity constraint. If $L_1 < L_2$, the invested dollar amount in the stock market under the constraint $X_t \le L_1$ is bounded by the corresponding money invested in the stock market for the level $L_2$. While an increasing level of $L$ invests in the portfolio's expected return, the portfolio becomes riskier. Therefore, a suitable level of $L$ depends on its {\em counter-effect} to the expected return and risk. 

Given the relationship between $W^*$ and $L$ in Proposition  \ref{pr:solution} and Proposition \ref{prop:solution}, a plausible method to set the capacity $L$ is to first estimate the number $W^*$ and solve the capacity $L$ conversely. For example, by estimating all expected costs in the retirement period, the retiree might be able to estimate $W^*$, say, 1 million. The Equation (\ref{eq:W*}) reduces one equation of the variable - the capacity $L$, which can be solved numerically.  In this way, Proposition \ref{pr:solution} and Proposition \ref{prop:solution} provide the optimal strategy when the wealth does not meet the threshold $W^*$ yet.

Finally, we demonstrate that (in Proposition \ref{pr:solution}) the optimal consumption rate is not a simple linear function of the wealth. Whether the wealth is greater than the threshold $W^*$ or not, the optimal consumption rate is a highly nonlinear function of the wealth. Therefore, standard annuities 
are not optimal from an optimal portfolio choice perspective, if the risk of living standard is a concern.

\section{Conclusion}

This paper solves an optimal portfolio choice problem under risk capacity constraint in an infinite horizon framework. We present an explicit consumption-saving policy for two critical situations. Then we apply our results to the asset allocation problem for a retiree with longevity risk and living standard risk when the retiree has a preference on a stream of consumption rates or inheritance wealth, respectively. We demonstrate that the risk capacity constraint implies a buffer-stock saving strategy and reduces the living standard risk. By contrast, the leverage constraint generates substantial living standard risk. Our discussions rely on the explicit characterization of the regions on which the risk capacity constraint is binding and a detailed analysis of the smooth property of the value function. 
%

\newpage
\renewcommand {\theequation}{A-\arabic{equation}} \setcounter
{equation}{0}

\section*{Appendix A. Proofs}
{\bf Proof of Proposition \ref{pr:general characterization}.}

 Zariphopoulou (1994) demonstrates the result for $\beta = 0$. We assume that $\beta>0$. It is standard to show that $\bar{V}(0)=\frac{\alpha+\beta}{ \delta}u(0)$ and $\bar{V}(W)$ is (strictly) continuous, increasing and concave.  We show that $\bar{V}(x)$ is the viscosity solution of (\ref{eq:hjb-general}) and such a solution is unique. 
 
 The existence part is standard in the theory of viscosity solution. See Fleming and Soner (2006, Chapter 3). 
To prove the uniqueness part it suffices to prove the following comparison principle: if $V_{1}(W)$ is the viscosity supersolution and $V_{2}(W)$ is the viscosity subsolution and satisfies $V_{1}(0) \geq V_{2}(0)$, then $V_{1}(W) \geq V_{2}(W)$ for all $W \in (0,\infty)$.

Since the function $u(W)$ is not Lipschitz, we cannot apply the standard comparison principle directly in our situation. For this purpose, we separate $(0,\infty)$ into two parts: $(0,\delta)$ and $(\delta,\infty)$ for a proper positive number $\delta$, then show that $\forall \epsilon>0$,$
V_{1}(W)+\epsilon \geq V_{2}(W), \ \ \forall W>0.$ 
Since $V_{1}(0) \geq V_{2}(0)$, there exists $\delta>0$, such that 
\be
\label{comp1}
V_{1}(W)+\epsilon \geq V_{2}(W), \ \ \forall W\in (0,\delta].
\ee
On the region $W\in (\delta,\infty)$, $u(W)$ is Lipchitz. Since $\psi(W)+\epsilon$ is the test function for $V_{1}(W)+\epsilon$, $V_{1}(W)$ is also a supersolution of (\ref{eq:hjb-general}), then we utilize the standard comparison principle in Fleming and Soner (2006, Chapter 5) to obtain
\be
\label{comp2}
V_{1}(W)+\epsilon \geq V_{2}(W), \ \ \forall  W \in (\delta,\infty)
\ee
Now, combine (\ref{comp1}) and (\ref{comp2}), we have 
\beq
V_{1}(W)+\epsilon \geq V_{2}(W), \ \ \forall W>0.
\eeq
Since $\epsilon$ is arbitrary, the comparison principle holds and the proof is now complete. 
\hfill$\Box$

{\bf Proof of Proposition \ref{pr:general without constraint}.} 

We prove Case (a), namely,  $\alpha>0$ and $\beta>0$.  Case (b) and Case (c) can be proved similarly.  

We assume the solution is in the form of 
\beq
\bar{V}(W)={A \over 1-R}W^{1-R}
\eeq
and plug it into the HJB equation:
\begin{equation}
\label{eq:hjb-general without}
\delta \bar{V}(W) = \max_{ X\geq 0} \left[ \mu X \bar{V}'(W) + \frac{1}{2} \sigma^2 X^2 \bar{V}''(W)  \right]  \\
 + \max_{c \ge 0} \{\alpha u(c) - c \bar{V}'(W)\}+\beta u(W),   (W > 0).
\end{equation}
 We obtain $c^*=({\alpha\over A})^{1\over R}W$, and Equation (\ref{eq:hjb-general without}) becomes:
\beq
{\delta A \over 1-R} W^{1-R}={\kappa A \over R}W^{1-R}+{\alpha\over 1-R}({\alpha\over A})^{{1\over R}-1}W^{1-R}-({\alpha\over A})^{1\over R} AW^{1-R}+{\beta\over 1-R}W^{1-R}
\eeq
By simplifying the above equation, we obtain
\be
\label{positive}
({\delta \over 1-R}-{\kappa\over R})A={\alpha}^{1\over R}{R\over 1-R} A^{1-{1\over R}}+{\beta\over 1-R}
\ee
It suffices to show that there is only one positive number $A$ which satisfy equation (\ref{positive}). Let 
\beq
h(t)=({\delta \over 1-R}-{\kappa\over R})t-{\alpha}^{1\over R}{R\over 1-R}t^{1-{1\over R}}-{\beta\over 1-R}.
\eeq
Then 
\beq
h'(t)=({\delta \over 1-R}-{\kappa\over R})+{\alpha}^{1\over R}t^{-{1\over R}},
\eeq
and 
\beq
h''(t)=-{1\over R} \alpha^{1\over R}t^{-1-{1\over R}}
\eeq

{\bf Case 1:} If $R<1$,  by Assumption A,  $h'(t)>0,\forall t\geq 0$. Moreover,  $h(0)=-\infty$ and $h(+\infty)=+\infty$. Since $h(\cdot)$ is a increasing function, there exists a unique positive number $A$ such that $h(A)=0$,  which satisfies (\ref{positive}).

{\bf Case 2:} If $R>1$, then $h(0)=-{\beta\over 1-R}>0$, $h(+\infty)=-\infty$. Moreover,  $h''(t)<0, \forall t>0$ and $h'(0)=+\infty$, $h'(+\infty)={\delta \over 1-R}-{\kappa\over R}<0$. Then $h(\cdot)$ is concave, increase first, and then decrease on $(0,\infty)$. Using the sign of $h(0)$ and $h(\infty)$, there is a unique positive number $A$ such that $h(A)=0$.
The proof is complete.  \hfill$\Box$

{\bf Proof of Proposition \ref{pr:region}.}

The ordinary differential equation for $\bar{V}(W)$ in the unconstrained and constrained region becomes,
\be
\label{eq:V-U1}
\delta \bar{V}(W) = \alpha^{1/R} \frac{R}{1-R}(\bar{V}')^{-(1-R)/R} + \kappa \frac{ (\bar{V}'(W)^2}{-\bar{V}''(W)}+\beta u(W),
\ee 
and
\be
\label{eq:V-B}
\delta \bar{V}(W) = \alpha^{1/R} \frac{R}{1-R}(\bar{V}')^{-(1-R)/R} + \mu L \bar{V}'(W) + \frac{1}{2} \sigma^2 L^2 \bar{V}''(W)+\beta u(W),
\ee
respectively. Define a function
\be
Y(W) = \mu \bar{V}'(W) + \sigma^2 L \bar{V}''(W), W > 0.
\ee
Then, $Y(W) < 0, \forall W \in {\cal U}$, and $Y(W) > 0$ for any $W \in {\cal B}$.

{\bf Step 1.} In the unconstrained region, the value function $\bar{V}(\cdot)$ satisfies the ODE (\ref{eq:V-U1}).
By differentiating the ODE equation once and twice, we obtain
\beq
\delta \bar{V}' = -\alpha^{1/R} (\bar{V}')^{-{1 \over R}}\bar{V}'' - 2 \kappa \bar{V}' + \frac{\kappa (\bar{V}')^2 \bar{V}'''}{(\bar{V}'')^2}+\beta u'(W)
\eeq
and
\begin{eqnarray*}
	\delta \bar{V}'' &=&  \alpha^{1/R} {1\over R}(\bar{V}')^{-{1\over R}-1}(\bar{V}'')^2- \alpha^{1/R} (\bar{V}')^{-{1\over R}}\bar{V}''' \\
	&& - 2 \kappa \bar{V}'' + \frac{\kappa (\bar{V}')^2 \bar{V}''''}{(\bar{V}'')^2} + \frac{2 \kappa \bar{V}' \bar{V}'''}{(\bar{V}'')^3} \left\{ (\bar{V}'')^2 - \bar{V}' \bar{V}''' \right\}+\beta u''(W)
\end{eqnarray*}
By the definition of $Y(W)$, the last two equations imply
\beq
\delta Y &=& \alpha^{1/R} (\bar{V}')^{-{1\over R}}\left[-Y'-{2\mu \over \sigma^2LR}Y+{1\over \sigma^2 LR}{Y^2\over \bar{V}'}+{\mu^2\over \sigma^2LR}\bar{V}'\right]- 2 \kappa Y \\
&&+ \frac{\kappa (\bar{V}')^2}{(\bar{V}'')^2} Y'' + \frac{2 \kappa \bar{V}' \bar{V}'''}{(\bar{V}'')^3} \left\{ \frac{\bar{V}''}{\sigma^2 L} Y - \frac{\bar{V}'}{\sigma^2 L} Y' \right\} \\
&& +\beta \left[ \mu u'(W)+L\sigma^2 u''(W)\right]
\eeq
We then define an elliptic operator on the unconstrained region by
\begin{eqnarray*}
	{\cal L}^{ {\cal U}}[y] &\equiv & - \frac{\kappa (\bar{V}')^2}{(\bar{V}'')^2} y'' -  \frac{2 \kappa \bar{V}' \bar{V}'''}{(\bar{V}'')^3} \left\{ \frac{\bar{V}''}{\sigma^2 L} y - \frac{\bar{V}'}{\sigma^2 L} y' \right\}  + ( \delta  + 2\kappa)y \\
	&& -\alpha^{1/R} (\bar{V}')^{-{1\over R}}\left[-y'-{2\mu \over \sigma^2LR}y+{1\over \sigma^2 LR}{y^2\over \bar{V}'}+{\mu^2\over \sigma^2LR}\bar{V}'\right] \\
	&&- \beta \left[ \mu u'(W)+L\sigma^2 u''(W)\right].
\end{eqnarray*}
Therefore, ${\cal L}^{ {\cal U} }[Y] =  0 $ in ${\cal U}$. 

{\bf Step 2. }  In the constrained region ${\cal B}$, by differentiating the ODE (\ref{eq:V-B}) of $V(W)$ once and twice, we have
\beq
\delta \bar{V}' = -\alpha^{1/R} (\bar{V}')^{-{1 \over R}}\bar{V}''  + \mu L \bar{V}'' + \frac{1}{2} \sigma^2 L^2 \bar{V}'''+\beta u'(W)
\eeq
and
\beq
\delta \bar{V}'' =  \alpha^{1/R} {1\over R}(\bar{V}')^{-{1\over R}-1}(\bar{V}'')^2-\alpha^{1/R} (\bar{V}')^{-{1\over R}}\bar{V}'''+ \mu L \bar{V}''' + \frac{1}{2} \sigma^2 L^2 \bar{V}''''+\beta u''(W).
\eeq
Then, 
\beq
\delta Y &=& \alpha^{1/R} (\bar{V}')^{-{1\over R}}\left[-Y'-{2\mu \over \sigma^2LR}Y+{1\over \sigma^2 LR}{Y^2\over \bar{V}'}+{\mu^2\over \sigma^2LR}\bar{V}'\right]+ \mu L Y' + \frac{1}{2} \sigma^2 L^2 Y'' \\
&& + \beta\left[ \mu u'(W) + \sigma^2 L u''(W) \right].
\eeq
Similarly, we define an elliptic operator
\beq
{\cal L}^{{\cal B}}[y] &=& - \frac{1}{2} \sigma^2 L^2 y'' - \mu L y' +    \delta y - \beta \mu u'(W) - \beta \sigma^2 L u''(W)\\
&-& \alpha^{1/r} (\bar{V}')^{-{1\over R}}\left[-y'-{2\mu \over \sigma^2LR}y+{1\over \sigma^2 LR}{y^2\over \bar{V}'}+{\mu^2\over \sigma^2LR}\bar{V}'\right]
\eeq
Then $ {\cal L}^{{\cal B}}[Y] = 0$ in ${\cal B}.$

{\bf Step 3.}  By straightforward calculation, we obtain
\be
\label{question}
{\cal L}^{ {\cal B}}[0]={\cal L}^{ {\cal U}}[0]=- \beta W^{-R-1} ( \mu W - \sigma^2 L R)- \alpha^{1/R} {\mu^2\over \sigma^2LR}(\bar{V}')^{1-{1\over R}}.
\ee
For simplicity, let
\beq
g(W) \equiv - \beta W^{-R-1} ( \mu W - \sigma^2 L R)- \alpha^{1/R} {\mu^2\over \sigma^2LR}(\bar{V}')^{1-{1\over R}}.
\eeq

{\bf Step 4.}  By the nontrivial  unconstrained region condition (1), there exists a real number $W_1 > 0$ such that $(0, W_1) \subseteq {\cal U}$ and $Y(W_1) = 0$. The existence of $W_1$ follows from Proposition \ref{pr:general without constraint} that ${\cal U} \ne (0, \infty)$. We show that $(W_1, \infty) \subseteq {\cal B}$ by a contradiction argument. 

Assume that, there exists $W_2 > W_1$ such that $(W_1, W_2) \subseteq {\cal B}$ and $Y(W_2) = 0$. Moreover, there exists $W_3 > W_2$ such that $(W_2, W_3) \subseteq {\cal U}$. We show this is impossible and thus finish the proof.

We first show that the constant function $y=0$ is {\em not} the supersolution for ${\cal L}^{{\cal B}}[y ]= 0$ in the region $(W_1, W_2)$. The reason is as follows. Otherwise, since ${\cal L}^{{\cal B}}[Y] = 0$ in the region $(W_1, W_2) \subseteq {\cal B}$ and $Y(W_1) = Y(W_2) = 0$, then by the comparison principle, $Y(W) \le y=0$ for $W \in (W_1, W_2)$. However, by its definition of ${\cal B}, Y(W) > 0$ for all $W \in (W_1, W_2)$. This contradiction show that the constant funciton $y=0$ is not a supersolution of ${\cal L}^{{\cal B}}[y ]= 0$ in the region $(W_1, W_2)$.
Therefore, there exists some $W_0 \in (W_1, W_2)$ such that, at $W = W_0$,
\beq
{\cal L}^{{\cal B}}[0 ] = g(W_0) < 0.
\eeq

We divide the proof into two situations because of the single crossing condition (3).

{\bf Case 1.}  The function $g(W)$ does not change sign at all in $(0, \infty)$.

In this case, $g(W) < 0$ for all $(0,\infty)$. We now consider the region $(W_2, W_3) \in {\cal U}$ and the operator ${\cal L}^{{\cal U}}$. Since ${\cal L}^{{\cal U}}[0] \le 0$ in this small region, the constant function $y=0$ is the subsolution for ${\cal L}^{{\cal U}}[0 ] = 0$. Since  $Y(W_2) = Y(W_3) = 0$, by the comparison principle, we obtain $Y(W) \ge 0, \forall W \in (W_2, W_3)$, which is impossible since $Y(W)$ is strictly negative over the region $(W_2, W_3) \subseteq {\cal U}$, the unconstrained region. 

{\bf Case 2.}  The function $g(W)$ change the sign in exactly one time.

In this case, $\beta$ must be positive. Otherwise, $g(W) = -\alpha^{1/R} {\mu^2\over \sigma^2LR}(\bar{V}')^{1-{1\over R}} < 0$ never change the sign.
By using the order of value function condition (2), $\bar{V}'(W)$ has the same order of $W^{-R}$, then $(\bar{V}')^{1-{1\over R}}$ has the same order of $W^{1-R}$. By comparing the order of $W$ of each term in in the function $g(W)$, the term $\sigma^2 LR\beta W^{-1-R}$ dominates other terms for small value of $W$. Then $
\lim_{W \downarrow 0} g(W) > 0$ (including positive infinite). Therefore, the function $g(W)$ must be negative for all $W > W_0$. In particular, $g(W) < 0, \forall W \in (W_2, W_3)$. Following the same proof as in Case 1, the constant $y=0$ is the subsolution for ${\cal L}^{{\cal U}}[0 ] = 0$. It implies that $Y(W) \ge 0, \forall W \in (W_2, W_3)$. This leads a contradiction again by the definition of ${\cal U}$.

By the above proof, we have shown that $(W_1, \infty) = {\cal B}$ by a contradiction argument.  \hfill$\Box$

\begin{lem}
\label{lem:appendix1}
In Problem (A), there exists $W^* > 0$ such that the open interval $(0,W^*)$ is included in the unconstrained $\tilde{\cal U}$ and $X^*(W^*)=L$
\end{lem}
\proof Assume not, then there exists a sequence of $W_n \rightarrow 0$, such that $X^{*}(W_n)=L$ and (from the definition of the constrained domain and its corresponding HJB equation):
\beq
\delta U(W_n) \geq {R \over 1-R}(U'(W_n))^{1-{1\over R}}+{1\over 2}\mu L U'(W_n).
\eeq
Set $f(t)={R \over 1-R}t^{1-{1\over R}}+{1\over 2}\mu Lt$. Then the function $f(t)$ attains its minimum at $t^*=({1\over 2}L\mu)^{-R}$ and $f(t^*)={1\over 1-R}[{1\over 2}L\mu]^{1-R}$.
We divide the proof in two cases.

{\bf Case 1.}  If $R<1$, let $n\rightarrow \infty$, then $\delta U(W_n)$ converges to $u(0) = 0$. On the other hand, the right side of the last inequality is bounded from below by a positive number ${1\over 1-R}[{1\over 2}L\mu]^{1-R}$, which is contradiction. 

{\bf Case 2.} If $R>1$, let $n \rightarrow \infty$, the $\delta U(W_n)$ tends to $-\infty$ but right side is always bounded below by a finite number, ${1\over 1-R}[{1\over 2}L\mu]^{1-R}$, another contradiction. We have thus finished the proof. \hfill$\Box$

{\bf Proof of Proposition \ref{pr:characterization2}.}

We prove Proposition \ref{pr:characterization2} by verifying condition (1) and (2) in  Proposition \ref{prop:region}, since the condition (3) holds naturally. Lemma \ref{lem:appendix1} implies condition (1) for the unconstrained region. As for condition (2), we first notice that $U(W) $ is bounded by the value function (Merton, 1971) without the risk constrain region. Therefore, there exists a positive number $C_1$ such that $U(W) \le C_1 W^{1-R}$. Moreover, by following a similar argument in Vila and Zariphoulou (1997, Lemma C.1, (ii)), it can be shown that $U(W) \ge C_0 W^{1-R}, \forall W \in (0, \infty)$ for a positive number $C_0$. Then, $U'(W)$ has the same order of $W^{-R}$. Alternatively, a similar argument in the following Lemma \ref{lem:derivative2} also demonstrates the order of value function condition (2) in this case. The proof is finished. \hfill$\Box$

\begin{rem}
 Vila and Zariphopoulou (1997, Proposition 4.4) shows a similar result under the constraint that $X_t \le k(W_t + L)$. We can modify some arguments in  Vila and Zariphopoulou (1997) to prove Proposition \ref{pr:characterization2} under the risk capacity constraint (the details are available upon request). However, this method cannot be used in Problem (B) and the general problem (\ref{value:general})).
\end{rem}

{\bf Proof of Proposition \ref{pr:solution}.}


By Karatzas and Shreve (1998), the value function in the unconstrained region can be written as  $J^{(m,n)}(W)$ for parameters $m, n$ as follows. For any real number $m, n$, and $W^* \ge 0$, we define the a class of strictly increasing and concave function,
$W \in (0, W^*] \rightarrow J^{(m,n)}(W) \in \mathbb{R}^{+}$ such that
\beq
W = W(C) \equiv \frac{1}{\lambda^{\infty}} \left(C + mC^{\omega^{+}} + n C^{\omega^{-}} \right),
\eeq
which is strictly increasing,
and
\beq
J(C) \equiv \frac{1}{\lambda^{\infty}} \left\{ \frac{C^{1-R}}{1-R} + m\frac{\omega^{+}}{\omega^{+} - R} C^{\omega^{+} - R}  + n \frac{\omega^{-}}{\omega^{-} - R} C^{\omega^{-} - R} \right\}.
\eeq
These two functions $W(C), J(C)$ with a variable $C$ clearly introduce a well-defined increasing and $C^2$ smooth function, and $U(W) = J(C)$.  It can be shown that when $W = 0$, the optimal consumption rate is zero. It implies that $n = 0$. Then $W = W(C) = H(C)$, and $W^* = H(C^*)$ for a unique finite number $C^*$. 
%
  It remains to solve $\{C^*, m \}$, by using both the value-matching and smooth-fit condition below.
 
%
%

We start with the smooth fit condition at $W^*$. The value function $U(W)$ satisfies
\be
L = -\frac{\mu}{\sigma^2} \frac{U_W}{U_{WW}}|_{W = W^*}
\ee
In the unconstrained region, noting that  $U_W = C^{-R}$, then $U_{W}(W^*) = (C^*)^{-R}$, and 
\begin{eqnarray*}
U_{WW}= -R C^{-R-1} \frac{1}{dW/dC} = -R\frac{1}{C^{R+1} H'(C)},
\end{eqnarray*}
then, we obtain 
\be
\label{eq:boundary-matching}
L = \frac{\mu}{\sigma^2 R} C^* H'(C^*).
\ee

We next consider the value function in the constrained region $(W^*, \infty)$ which satisfies
\be
\frac{R}{1-R} U_{W}^{1 - \frac{1}{R}} + \mu L U_{W} + \frac{1}{2}\sigma^2 L^2 U_{WW} - \delta U = 0.
\ee
We define a function $K(x):[W^*, \infty) \rightarrow (0, \infty)$ by the ordinary differential equation
\be
\frac{1}{2} \sigma^2 L^2 K''(x) = \delta K(x) - \frac{R}{1-R} K'(x)^{1-\frac{1}{R}} - \mu L K'(x).
\ee
The value-matching condition is 
\be
K(W^*) = J(C^*),
\ee
 and the smooth-fit condition at $W^*$ is
\be
\label{eq:first-order}
K'(W^*) = (C^*)^{-R}.
\ee
It is not sufficient to characterize the function $K(x)$ yet as $W^*$ is to be solved. For the end, we notice that $U(W,0) \le U(W) \le U^{\infty}(W)$, $K(W)$ has the order as $W^{1-R}$ when $W \rightarrow \infty$. 

{\bf Case 1.} If $R > 1$, then $\lim_{W \rightarrow \infty} U(W) =0$. Then  we derive another boundary condition $K(\infty) = 0$.

If this case, the ordinary differential equation theory (King, Billingham and Otto, 2003) characterizes the function $K(x)$ and $W^*$. Specifically, we define a function $h: [0, \frac{1}{W^*} ] \rightarrow (0, \infty), h(W) =  K(\frac{1}{W})$. Then $\lim_{W\rightarrow 0}h(W) = 0, h(\frac{1}{W^*}) = J(C^*)$, and it is straightforward to see that $h(x)$ satisfies a second-order differential equation $h''(x) = H(x, h(x), h'(x))$.
Finally, $W^*$ satisfies the following equation (since $K'(W^*) = (C^*)^{-R}$)
\begin{eqnarray}
(C^*)^{-R} = h'(\frac{1}{W^*} ) \frac{-1}{ (W^{*})^2}, 
\end{eqnarray}
implying 
\be
 h'(\frac{1}{W^*})  = -(W^{*})^2 (C^*)^{-R} = - H(C^*)^2 (C^*)^{-R}.
 \ee

{\bf Case 2.} If $R < 1$, then $\lim_{W \rightarrow \infty} U(W) = \infty$. 

We define another function $k(x):[0, \frac{1}{W}] \rightarrow (0, \infty)$ by $k(x) = \frac{1}{K(1/x)}$.
Then $\lim_{W \rightarrow 0}k(W) = 0, k(\frac{1}{W^*}) = \frac{1}{J(C^*)}$. It is straightforward to verify that $k(x)$ satisfies a second-order ordinary differential equation.  Moreover,
\beq
k'(x) = \frac{1}{K(1/x)^2} K'(1/x)\frac{1}{x^2},
\eeq
implying
\be
k'(\frac{1}{W^*}) = \frac{(C^*)^{-R}}{J(C^*)^{2}}H(C^*)^{2}.
\ee

By using the characterization of the value function in Proposition \ref{pr:general characterization} and the value function is $C^2$ smooth, the existence and the uniqueness of these two parameters ${C^*,m}$ is guaranteed by two equations (\ref{eq:boundary-matching}) and (\ref{eq:first-order}). \hfill$\Box$

To simplify the notations, we use $V^{0}(W), V^{\infty}(W)$ to represent $V(W,0), V(W,\infty)$.

We prove two lemmas in proving Proposition \ref{prop:region}.

%

\begin{lem}
\label{lem:derivative2}
Assume $V(W)$ is $C^2$ smooth, then there exists two positive numbers $C_0, C_1$ such that \beq
C_{0} W^{-R} \le V'(W) \le C_{1} W^{-R}, \forall W > 0.
\eeq
In particular,  $\lim_{W \rightarrow 0} V'(W)=\infty $ and $\lim_{W \rightarrow \infty} V'(W)= 0$.
\end{lem}

\proof 
By a direct calculation,  $V^{0}(W) = \frac{u(W)}{ \delta }$ and Proposition \ref{pr:general without constraint} states that $V^{\infty}(W) = \frac{u(W)}{ \delta -  \rho}$. 
Then,  by using the concave property of the function $V(\cdot)$, for any positive number $W>0$ and $E>0$, we have 
\beq
V'(W)&\geq& {1 \over E}[V(W+E)-V(W)]\\
&\geq & {1\over E}[V^{0}(W+E)-V^{\infty}(W)]\\
&=& \frac{1}{E}\left[ \frac{1}{1-R} \frac{1}{ \delta}(W+E)^{1-R}- \frac{1}{1-R} \frac{1}{\delta - \rho}W^{1-R}\right].
\eeq
Choosing $E=kW$, we have
\beq
V'(W)\geq {1\over k}\left[\frac{1}{1-R} \frac{1}{ \delta}(k+1)^{1-R}-\frac{1}{1-R} \frac{1}{ \delta -\rho}\right]W^{-R}
\eeq
Let
\be
\label{eq:C0}
C_{0} = \sup_{k >0} \frac{1}{k(1-R)} \left[\frac{1}{\delta}(k+1)^{1-R}- \frac{1}{ \delta -\rho}\right]
\ee
where $x^{+} = \max(x, 0)$. It is easy to see $C_0$ is positive no matter $R>1$ or $R<1$.

By the same reason, for any $E = \beta W, \beta \in (0,1)$, we have
\begin{eqnarray*}
V'(W) & \le & \frac{1}{\beta W} \left[ V(W) - V(W - \beta W) \right] \\
& \le & \frac{1}{\beta W} \left[  V^{\infty}(W) - V^{0}(W - \beta W) \right] \\
&  \le & \frac{1}{\beta(1 - R)} \left\{  \frac{1}{\delta - \rho } - \frac{1}{ \delta } (1-\beta)^{1-R} \right\} W^{-R}.
\end{eqnarray*}
Let
\be
\label{eq:C1}
C_{1} = \inf_{0 < \beta < 1} \frac{1}{\beta(1 - R)} \left[  \frac{1}{ \delta - \rho } - \frac{1}{ \delta } (1-\beta)^{1-R} \right] \ee
Clearly,  $C_1$ is positive no matter $R>1$ or $R<1$. The proof is finished.
  \hfill$\Box$

\begin{rem}
	The proof of Lemma \ref{lem:derivative2} is similar to a method in Xu and Yi (2016) for an optimal portfolio choice problem on a consumption constraint.
\end{rem}
\begin{lem}
\label{lem:middle}
Assume $V(\cdot)$ is $C^2$ smooth, then there exists $\tilde{W}$ such that the open interval $(0,\tilde{W})$ is included in ${\cal U}$, and $X^*(\tilde W)=L.$
\end{lem}
\proof Assume not, then there exists a sequence $W_n\rightarrow 0$ such that $X^*(W_n)=L$. By using equation (\ref{eq:BB}) and Lemma \ref{lem:derivative2},  we have
\begin{eqnarray*}
\label{Lemma 2.5}
\delta V(W_n)  &\geq &   {1\over 1-R} W_n^{1-R} +{1\over 2} \mu LV'(W_n) \\
&\geq & {1\over 1-R} W_n^{1-R}+{1\over 2}\mu L C_0 {W_n}^{-R}\\
&=& W_n^{-R} ({1\over 1-R} W_n+{1\over 2}\mu LC_0)
\end{eqnarray*} 

{\bf Case 1. $R < 1$.} Since $V(W)$ is continuous , as $n \rightarrow \infty$,  the left hand side of the last inequality  approaches to $\delta V(0) = 0$.

{\bf Case 2. $R>1$.} The left hand side of the last inequality approaches to $\delta V(0) = -\infty$.  However,  the term $W_n^{-R} ({1\over 1-R} W_n+{1\over 2}\mu LC_0)$ tends to $+\infty$ on the right hand side of the last equality, which is a contradiction. \hfill$\Box$

{\bf Proof of Proposition \ref{prop:region}.} 

Since the single crossing condition holds for the function $g(W)  = - [\mu u'(W) + \sigma^2 L u''(W)] = -W^{-R-1} (\mu W - \sigma^2  L R) $ in this case, this proposition follows from Proposition \ref{pr:region}, Lemma \ref{lem:derivative2} and Lemma \ref{lem:middle}. \hfill$\Box$

%

\begin{lem}
\label{lem:smooth-fit}
Let $F: (0, \infty)  \times \mathbb{R} \times \mathbb{R} \times \mathbb{R} \rightarrow \mathbb{R}$ be a continuous and elliptic operator, that is,
 $F(x,r,p, X) \le F(x,r,p, Y), \forall X \ge Y$. 
Assume $V(x)$ is a continuous viscosity solution of a second-order (HJB) equation $F(x, u, u_x, u_{xx}) = 0$ and the region of $x$ is ${\cal D} = (0, \infty)$. Moreover, there exists $x^*$ such that $V(x)$ is smooth in both $(0, x^*)$ and $(x^*, \infty)$, then $V(x)$ must satisfies the smooth-fit condition at $x^*$, that is, $V'(x^*-) = V'(x^*+)$.
\end{lem}

\proof Without lost of generality, we assume that $V'(x^{*}-) < 0 < V'(x^{*}+)$ and derive a contradiction. Since there is no available test function, the subsolution holds automatically.  We next check the supersolution. Let the test function in the form of 
\beq
\psi(x) \equiv V(x^*) + \frac{1}{2} \left[V'(x^{*}-) + V'(x^{*}+)  \right](x-x^*) + \alpha(x-x^*)^2
\eeq
We claim that $\alpha$ can take any real value: To make $\psi(x)$ the valid test function, we need to guarantee that $ \psi(x)\leq V(x)$ when $x$ is in a small neighborhood of $x^*$.  However, when $x\rightarrow x^*$, the linear term $\frac{1}{2} \left[V'(x^{*}-) + V'(x^{*}+)  \right](x-x^*)$ will dominate the quadratic term $\alpha(x-x^*)^2$. Therefore, when $x$ and $x^*$ are close enough, we could choose sufficiently large $\alpha$ such that $\psi(x)\leq V(x)$. It is now clear that $\alpha$ can take any value. \\
Now, apply the viscosity property at $x^*$, we have
\beq
F\left(x^*, V(x^*), \frac{1}{2} \left[V'(x^{*}-) + V'(x^{*}+)  \right], 2 \alpha\right) \ge 0,
\eeq
which is impossible by the free choice of the parameter $\alpha$.  \hfill$\Box$

\begin{rem}
Lemma \ref{lem:smooth-fit} can be viewed as a converse statement of Proposition \ref{prop:region}. 
If the value function is smooth in each region $(0, W^*), (W^*, \infty)$, then the value function must be smooth as long as the value function is continuous and a viscosity solution of a HJB equation. 
\end{rem}

{\bf Proof of Proposition \ref{prop:solution}.}


We divide the proof into several steps.

%
%
{\bf Step 1.}
Assuming $W^*$ is known, we derive candidate solution of Equation (\ref{eq:V-B}) in the constrained region. To simplify notation we still use $V(W)$ to represent the feasible solution of the value function, being a solution of a corresponding ODE.


The solution of the homogeneous ODE, $\frac{1}{2} \sigma^2 L^2 V_{WW} + L \mu V_{W} -  \delta V(W) = 0$, is  written as 
 $C_1 e^{\beta_1 W} + C_2 e^{\beta_2 W}.$
By the method of partial integral, one particular solution for the non-linear ODE (\ref{eq:V-B}) is
\be
V_{0}(W) = - \int_{0}^{W}\frac{2}{\sigma^2 L^2}u(x) 
\left\{ \frac{e^{\beta_1 x} e^{\beta_2 W} - 
e^{\beta_1 W} e^{\beta_2 x}  }{ W(e^{\beta_1 x}, 
e^{\beta_2 x}) } \right\} dx
\ee
where $W(f,g) = fg' - f'g$ is the Wronskian determinants of two solutions $\{f, g\}$ of a homogeneous second-order ODE. By a straightforward calculation, 
\begin{eqnarray*}
V_{0}(W) & = & \frac{2}{(\beta_1 - \beta_2)(1-R) \sigma^2 L^2} \\
&& \times  \left\{ e^{\beta_2 W} \int_{0}^{W} x^{1-R} e^{-\beta_2 x}dx -
e^{\beta_1 W } \int_{0}^{W} x^{1-R} e^{-\beta_1 x}dx \right\}.
\end{eqnarray*}
Therefore, the function $V_{0}(W)$ is well-defined and it can be expressed in terms of the incomplete gamma function.
A general solution of the ODE (\ref{eq:V-B}) is
\be
\label{eq:C}
V(W) = C_1 e^{\beta_1 W} + C_2 e^{\beta_2 W} + V_{0}(W).
\ee

{\bf Step 2.} Assuming $W^*$ is known, we show that $C_1={2\over (\beta_1-\beta_2)(1-R)\sigma^2L^2}(\beta_1)^{R-2}\Gamma(2-R)$  in Equation (\ref{eq:C}). 

 By Proposition \ref{pr:general without constraint}, $\frac{V(W)}{W^{1-R}}$ is bounded above by a constant. Therefore, $V(W)/e^{\beta_1 W}\rightarrow 0$ as $W\rightarrow\infty$ in the constrained region. On the other hand, by (\ref{eq:C}), as $W\rightarrow \infty$
 \be \label{C_1}
 C_1+C_2e^{(\beta_2-\beta_1)W}+{V_0(W)\over e^{\beta_1 W}}\rightarrow 0
 \ee
 Note that
 \be
 \label{V_0 over}
\frac{V_{0}(W)}{e^{\beta_1 W}}  =  \frac{2}{(\beta_1 - \beta_2)(1-R) \sigma^2 L^2}  \times  \left\{ e^{(\beta_2-\beta_1) W} \int_{0}^{W} x^{1-R} e^{-\beta_2 x}dx - \int_{0}^{W} x^{1-R} e^{-\beta_1 x}dx \right\}.
 \ee
 For the the first term in the bracket of (\ref{V_0 over}), since $\beta_2 < 0$, we have
 \beq
 e^{(\beta_2-\beta_1) W} \int_{0}^{W} x^{1-R} e^{-\beta_2 x}dx&=&e^{-\beta_1 W}\int_0^W x^{1-R} e^{\beta_2(W-x)}dx\\
 &\leq& e^{-\beta_1 W}\int_0^W x^{1-R} dx\\
 &=&e^{-\beta_1 W}{W^{2-R}\over 2-R}
 \eeq
which tends to 0 as $W\rightarrow \infty$.\\ For the second term in the bracket of (\ref{V_0 over}), change of variable $y=\beta_1 x$ leads to
\beq
\int_{0}^{W} x^{1-R} e^{-\beta_1 x}dx=(\beta_1)^{R-2}\int_0^{\beta_1 W}y^{1-R}e^{-y}dy.
\eeq
By the property of incomplete Gamma function (\ref{eq:Gamma1}) in Appendix B, 
\beq
(\beta_1)^{R-2}\int_0^{\beta_1 W}y^{1-R}e^{-y}dy \rightarrow (\beta_1)^{R-2}\Gamma(2-R).
\eeq
Then, we obtain
\beq
C_1={2\over (\beta_1-\beta_2)(1-R)\sigma^2L^2}(\beta_1)^{R-2}\Gamma(2-R).
\eeq
 In Step 5 below, we show that $C_2 = C(W^*)$ in Equation (\ref{eq:C*}).

{\bf Step 3.} Assuming $W^*$ is known, we characterize the feasible solution in  the unconstrained region.

We introduce a new variable $g$ by $V'(W)=g^{-R}$. Since $V(\cdot)$ is concave by a standard argument, $V'(W)$ is a decreasing function. Then, $W=G(g)$ for an increasing function $G(\cdot)$. Similarly, we can write $g$ as a well-defined increasing function of $W$, $g = g(W)$.  We characterize the function $G(\cdot)$ and derive the feasible function in terms of the auxiliary function $G(\cdot)$.

Since $W = G(g(W))$, then $1 = G'(g) g'(W)$, yielding $G'(g) = 1/g'(W)$. By using $
V''(W)={-Rg^{-R-1}\over G'(g)},$ the HJB equation becomes 
\beq
 \delta V(G(g))={1\over 1-R}[G(g)]^{1-R} + {\kappa \over R}g^{-R+1}G'(g).
\eeq
We differentiate both sides of the above equation again with respect to $W$, obtaining
\be
\label{G}
G''(g)={R\over \kappa}g^{-1}G'(g)\left[ \delta - \rho -g^{R}(G(g))^{-R}\right]
\ee
Since $G(\cdot)$ is strictly increasing, the unconstrained region of $W$, $ W \le W^*$,  corresponds one-one to a region of $g$, $g \le g^* = g(W^*)$. Moreover, for any $W \le W^*$,
\begin{eqnarray*}
V(W) & = & \frac{u(0)}{ \delta} + \int_{0}^{W} V_{W} dW \\
& = & \frac{u(0)}{ \delta } + \int_{0}^{G^{-1}(W)} g^{-R} G'(g) dg.
\end{eqnarray*}
Therefore, the feasible value function in the unconstrained region is uniquely determined by the auxiliary function $G(\cdot)$. The number $g^*$ is shown to be $D(W^*)$ in Step 5 below.

{\bf Step 4.} Assuming $W^*$ is known, we derive the boundary condition for ordinary differential equation (\ref{G}). 

Since $V'(0)=+\infty$ (Lemma \ref{lem:derivative2}), we have $G(0)=0$. Second, at $W=W^*$, $G(g^*)=W^*$. Moreover, the constraint $-{\mu \over \sigma^2}{V'(W^*-)\over V''(W^*-)}=L$ implies that 
\beq
G'(g^*)={LR\sigma^2 \over \mu}(g^*)^{-1}.
\eeq
 By the characterization of the feasible value function in Step 3, the required smooth-fit condition is  \be
 \label{eq:g-star}
(g^*)^{-R} ={2\over (\beta_1-\beta_2)(1-R)\sigma^2L^2}(\beta_1)^{R-1}\Gamma(2-R)e^{\beta_1W}+ C_{2} \beta_2e^{\beta_2 W^*}+V_0'(W^*).
\ee
Therefore, the boundary condition of the ODE (\ref{G}) are  $G(0)=0$, $G(g^*) = W^*$ and $G'(g^*)={LR\sigma^2\over \mu}(g^*)^{-1}$. It remans to determine $g^*$ and $C$ in the next step. 

{\bf Step 5.} Assuming the smooth-fit condition of the value function $V(W)$,  we show that $C_2 = C(W^*)$ and $g^* = D(W^*)$. 

The smooth-fit condition can be written as  $-{\mu \over \sigma^2}{V'(W^*+)\over V''(W^*+)}=L$. Then, the feasible function in Step 2 implies that 
\beq
-\mu \left[{2\over (\beta_1-\beta_2)(1-R)\sigma^2L^2}(\beta_1)^{R-1}\Gamma(2-R)e^{\beta_1W}+C_2\beta_2 e^{\beta_2 W^*}+V_0'(W^*)\right]\\
=\sigma^2L \left[{2\over (\beta_1-\beta_2)(1-R)\sigma^2L^2}(\beta_1)^{R}\Gamma(2-R)e^{\beta_1W}+C_2\beta_2^2e^{\beta_2 W^*}+V_0''(W^*)\right]
\eeq
Solving this equation, we obtain $C_2 = C(W^*)$ as in (\ref{eq:C*}).  By Equation (\ref{eq:g-star}), we have $g^* = D(W^*)$.


{\bf Step 6.} Assuming the existence of $W^*$ (which is guaranteed under assumption of $C^2$ smooth condition of the value function), we derive the equation of the parameter $W^*$.

 In fact, the value-matching equation, $V(W^*-)=V(W^*+)$, can be written as 
\be
\label{Match}
\frac{u(0)}{ \delta}+\int_0^{g^*} g^{-R}G'(g)dg={2\over (\beta_1-\beta_2)(1-R)\sigma^2L^2}(\beta_1)^{R-2}\Gamma(2-R)e^{\beta_1W^*}+C(W^*) e^{\beta_2 W^*}+V_0(W^*).
\ee
This is a one-variable equation of the variable $W^*$, as $g^* = D(W^*)$. This equation is the same as Equation (\ref{eq:W*}) proposed in the statement of this proposition.

{\bf Step 7.} (Necessary condition)  If the value function is $C^2$ smooth, then by Proposition \ref{prop:region}, there is a positive number $W^*$ such that the unconstrained region and the constrained region are separated by this number $W^*$. By Step 1 - Step 6 this number $W^*$ must satisfy Equation (\ref{Match}) to ensure the smooth-fit condition of the value function. Moreover, the positive solution of Equation (\ref{Match}) must be unique by Proposition \ref{prop:region} again. By its construction, $V(W, L) =  \tilde{V}(W,L)$ by (\ref{eq:solution}). The necessary part is proved.

{\bf Step 8.} (Verification). We assume the existence of a positive solution $W^*$ of Equation (\ref{Match}) and show that the value function is $C^2$ smooth.  

In fact, by Step 1 - Step 6, the function $\tilde{V}(W,L)$ is $C^2$ smooth, and a smooth solution of the HJB equation in each region $(0, W^*)$ and $(W^*, \infty)$. It remains to show that $\tilde{V}(W,L)$ is a viscosity solution of the HJB equation.  We show that, $\frac{\mu \tilde{V}_{W}(W,L)}{-\sigma^2 \tilde{V}_{WW}(W,L)}  \le L, \forall W \le W^*$. By its definition, it suffices to show that the function $g G'(g)$ is increasing since $\frac{\mu \tilde{V}_{W}(W,L)}{-\sigma^2 \tilde{V}_{WW}(W,L)} = \frac{\mu}{R \sigma^2} g G'(g)$ and $\frac{\mu}{R \sigma^2} g^* G'(g^*)= L$. Here, the function $G(\cdot)$ is defined by $W^*$. For this purpose, we notice that
\begin{eqnarray*}
(g G'(g))'  &= &  gG''(g) + G'(g)  \\
& = & G'(g) \left[ \frac{R}{\kappa} ( \delta - \rho - g^{R} G^{-R} )\right] + G'(g)
\end{eqnarray*}
in which Equation (\ref{eq:G}) is used. Since $G'(g) > 0$ follows from the concavity of the function $V(W)$ in the region $W \le L$, it reduces to show that
\beq
\left[ \frac{R}{\kappa} ( \delta - \rho - g^{R} G^{-R}) \right] + 1 \ge 0,
\eeq
Or equivalently, $ g^{R} G^{-R} <   \delta - \rho + \frac{\kappa}{R}$, as proposed in the proposition. Therefore, we have proved that
\beq
\left\{ \frac{\mu}{-\sigma^2 } \frac{\tilde{V}_{W}(W,L)}{\tilde{V}_{WW}(W,L)} \le L \right\} = (0, W^*],
\eeq
and
\beq
\left\{ \frac{\mu}{-\sigma^2 } \frac{\tilde{V}_{W}(W,L)}{\tilde{V}_{WW}(W,L)} \ge L \right\} = [W^*, \infty).
\eeq
By its construction, $V(W,L) =  \tilde{V}(W, L)$ is $C^2$ smooth, and it is the value function by Proposition \ref{pr:general characterization}. Moreover, by Step 7 and Proposition \ref{prop:region}, the positive number $W^*$ must be the unique positive solution of Equation (\ref{Match}). We have thus proved the sufficient part (the verification theorem).
\hfill$\Box$

%
%

{\bf Proof of Proposition \ref{prop:alternative}.} 

By using the same argument in proving Proposition \ref{pr:general characterization}, we can prove that the value function is the unique viscosity solution of the HJB equation (for $V(W) = V^{b}(W)$)
\beq
\delta V(W)=\max_{0 \leq X\leq bW} \left[{1\over 2}\sigma^2X^2V''+ \mu XV'\right]+ \max_{c \ge 0} \left\{ \alpha u(c) - c V(W) \right\} + \beta u(W)
\eeq
with initial value $V(0)=0$. Similar to Proposition \ref{pr:general without constraint}, we  find a $C^2$ solution of the form $V(W)=B\frac{W^{1-R}}{1-R}$ to the above HJB equation for a positive number $B$.

By plugging $V(W)=B\frac{W^{1-R}}{1-R}$ into the HJB equation with $X^* = bW$, a straightforward computation implies that 
\begin{eqnarray*}
	\delta B\frac{W^{1-R}}{1-R}  &=&{1\over 2}\sigma^2 b^2W^2 B (-R)W^{-R-1}+ \mu bW B W^{-R}+\beta {1\over 1-R}W^{1-R} + \alpha^{1/R} \frac{R}{1-R} (B W^{-R)^{1-1/R}})
\end{eqnarray*}
yielding an equation of $B$ as follows
\be
\label{eq:B}
\left[\delta + (1-R) (\frac{1}{2} \sigma^2 b^2 R - \mu b)\right] B=\beta + \alpha^{1/R} R B^{1-1/R}. 
\ee
Since $b < \frac{\mu}{R \sigma^2}$, then $X^* = bW$ is the solution in $\max_{0 \le X \le bW} \left[\frac{1}{2} \sigma^2 X^2 V'' + \mu X V'  \right]$. Moreover, the optimal consumption rate satisfies that $\alpha u'(c) = V'(W) = B W^{-R}$.
It remains to show the unique positive solution $B$ of Equation (\ref{eq:B}) for $\alpha > 0, \beta > 0$. We notice that, since $0 \le b \le \frac{\mu}{R \sigma^2}$, we have
\be
-\frac{\kappa}{R} \le \frac{1}{2} \sigma^2 b^2 R - \mu b \le 0.
\ee

{\bf  Case 1.} If $R < 1$, then by Assumption A,
\beq
\delta + (1-R)(\frac{1}{2}\sigma^2 b^2 R - \mu b) \ge \delta - (1-R) \frac{\kappa}{R} > 0. 
\eeq
Moreover, the left side of Equation (\ref{eq:B}) increases while the right side decreases with $B$, the existence and uniqueness of a positive solution $B$ is evident.

{\bf Case 2.} If $R > 1$, the
\beq
\delta + (1-R)(\frac{1}{2}\sigma^2 b^2 R - \mu b) \ge \delta >0.
\eeq
In this situation, the right side is increasing and concave. Moreover, as $W \rightarrow \infty$, the right side is dominated by the left side of equation, a linear function of $B$. Therefore, it is straightforward to see the existence and uniqueness of the number $B$.
 The proof is completed.
\hfill$\Box$

{\bf Proof of Corollary \ref{prop:strategy}.} 

In the unconstrained region, $V_{W} = g^{-R}$. Since $V''(W) = \frac{-R g^{-R-1}}{G'(g)}$, the optimal strategy is $X(W) = \frac{\mu}{R \sigma^2} g G'(g)$. $G(\cdot)$ is not a linear function in general. Otherwise, $W^* = \frac{R \sigma^2}{\mu} L$. Equation (\ref{eq:W*}) is viewed as an equation of of $L$, in which both sides are analytical function of the variable $L$. By the analytical function property, it cannot hold for a general choice of the capacity level $L$. \hfill$\Box$

\newpage
\renewcommand {\theequation}{B-\arabic{equation}} \setcounter
{equation}{0}

\section*{Appendix B: Incomplete Gamma function}
The lower incomplete gamma function and the upper incomplete gamma function are defined by 
by
\be
\Gamma(s,x) = \int_{x}^{\infty} t^{s-1} e^{-t}dt; \gamma(s,x) = \int_{0}^{x} t^{s-1}e^{-t}dt.
\ee
For any $Re(s) > 0$, the functions  $\Gamma(s,x)$ and $\gamma(s,x)$ can be defined easily. Each of them can be developed into a holomorphic function. In fact, the incomplete Gamma function is well-defined for all complex $s$ and $x$,  by using the power series expansion
\be
\gamma(s,x) = x^{s} \Gamma(s) e^{-x} \sum_{k=0}^{\infty} \frac{x^{k}}{\Gamma(s+k+1)}.
\ee

The following  asymptotic behavior for the incomplete gamma function are used in the proof of Proposition \ref{prop:solution}.
\be
\label{eq:Gamma1}
\lim_{x \rightarrow \infty} \gamma(s,x) = \Gamma(s),
\ee
and
\be
\label{eq:Gamma2}
\lim_{x \rightarrow 0} \frac{\gamma(s,x)}{x^{s}} = \frac{1}{s}.
\ee
See  N.M. Temme, ``The asymptotic expansion of the incomplete gamma functions" , SIAM J. Math. Anal.  10 (1979),  pp. 757 - 766.
%

It can also be connected with Kummer's Confluent Hypergeometric Function, when $Re(z) > 0$,
\be
\gamma(s,z) = s^{-1} z^{s} e^{-z} M(1, s+1, z)
\ee
where
\be
M(1,s+1,z) = 1 + \frac{z}{(s+1)} + \frac{z^2}{(s+1)(s+2)} + ...
\ee
Therefore, the incomplete Gamma functions can be computed effectively.

					\clearpage
\vspace{10pt}
		
					\begin{figure}[!tp]
					\label{fig:G}
\begin{center}
					\includegraphics[width=.98\textwidth] {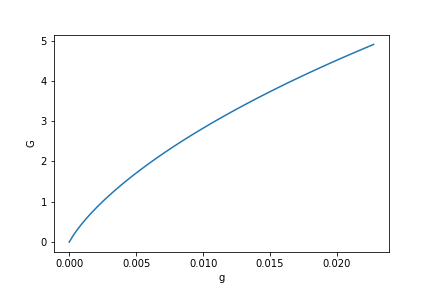}
\end{center}
					\caption{This figure displays the auxiliary function $G(g)$ in the unconstrained region in Proposition \ref{prop:solution} and Corollary \ref{prop:strategy}. The model parameters are $\mu = 0.1, \sigma = 0.3, R = 0.5, l = 0.7,$ and $W_0 = 1,000,000$. The x-axis represents the parameter $g$ and the y-axis represents $G(g)$ (in the unit of 100,000). As shown, this function is NOT a linear function, thus, the optimal strategy is not a myopic one as shown in Corollary \ref{prop:strategy}.  }
\end{figure}		

	\clearpage
\vspace{10pt}
		
					\begin{figure}[!tp]
					\label{fig:Dollar}
\begin{center}
					\includegraphics[width=.98\textwidth] {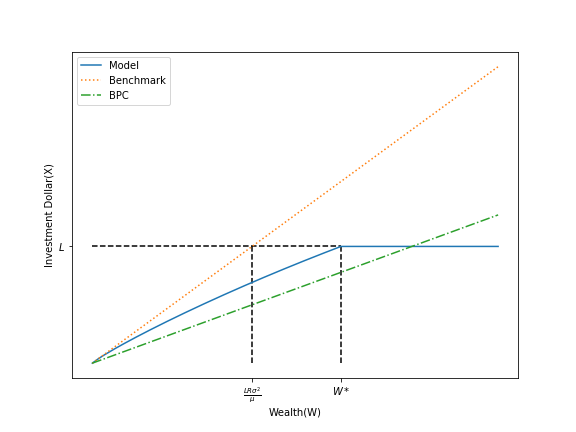}
\end{center}
					\caption{This figure displays the optimal portfolio strategy under three different strategies for $0 < R < 1$.  ``Model" denotes the model in Problem (B) under a risk capacity constraint $X_t  \le L = 0.7 W_0$. Parameters are $\mu = 0.1, \sigma = 0.3, R = 0.5, l = 0.7.$ By calculation, the wealth threshold $W^* = 490,235$ above which the retiree invests 700,000 in the stock market. When the wealth portfolio is smaller than $W^*$, the optimal strategy is $\frac{\mu}{R \sigma^2} g G'(g)$ where the auxiliary function $G(\cdot)$ is illustrated in Figure 1. ``Benchmark" denotes the optimal dollar amount in Proposition \ref{pr:general without constraint} in the absence of the constraint on the risky asset investment. ``BPC" denotes the optimal strategy (in Proposition \ref{prop:alternative}) under a leverage constraint that $X_t \le \frac{1}{2} \frac{\mu}{R \sigma^2} W_t$.}
\end{figure}

				\clearpage
\vspace{10pt}
		
					\begin{figure}[!tp]
					\label{fig:Percent}
\begin{center}
					\includegraphics[width=.98\textwidth] {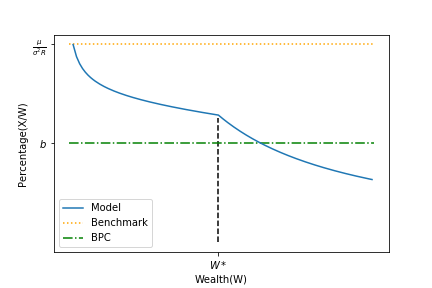}
\end{center}
					\caption{This figure displays the optimal percentage of wealth, $\frac{X(W)}{W}$, invested in the stock market in Problem (B). The parameters are the same as in Figure 2. ``BPC" denotes the optimal strategy (in Proposition \ref{prop:alternative}) under a leverage constraint that $X_t \le b W_t$ where $b=\frac{1}{2} \frac{\mu}{R \sigma^2}$. As shown, the percentage is decreasing in the entire region of $W$. We also notice that the percentage curve is steeper in the beginning of the retirement time when the wealth is closes to initial wealth than that when the wealth closes to the threshold $W^*$. As a function of $W$, $\frac{X(W)}{W}$ is not $C^1$ smooth under the risk capacity constraint,  in contrast to the standard model (Proposition \ref{pr:general without constraint}) or the model under leverage constraint (Proposition \ref{prop:alternative}). }
					 \end{figure}

				\clearpage
\vspace{10pt}

\begin{figure}[!tp]
					\label{fig:Dollar2B}
\begin{center}
					\includegraphics[width=.98\textwidth] {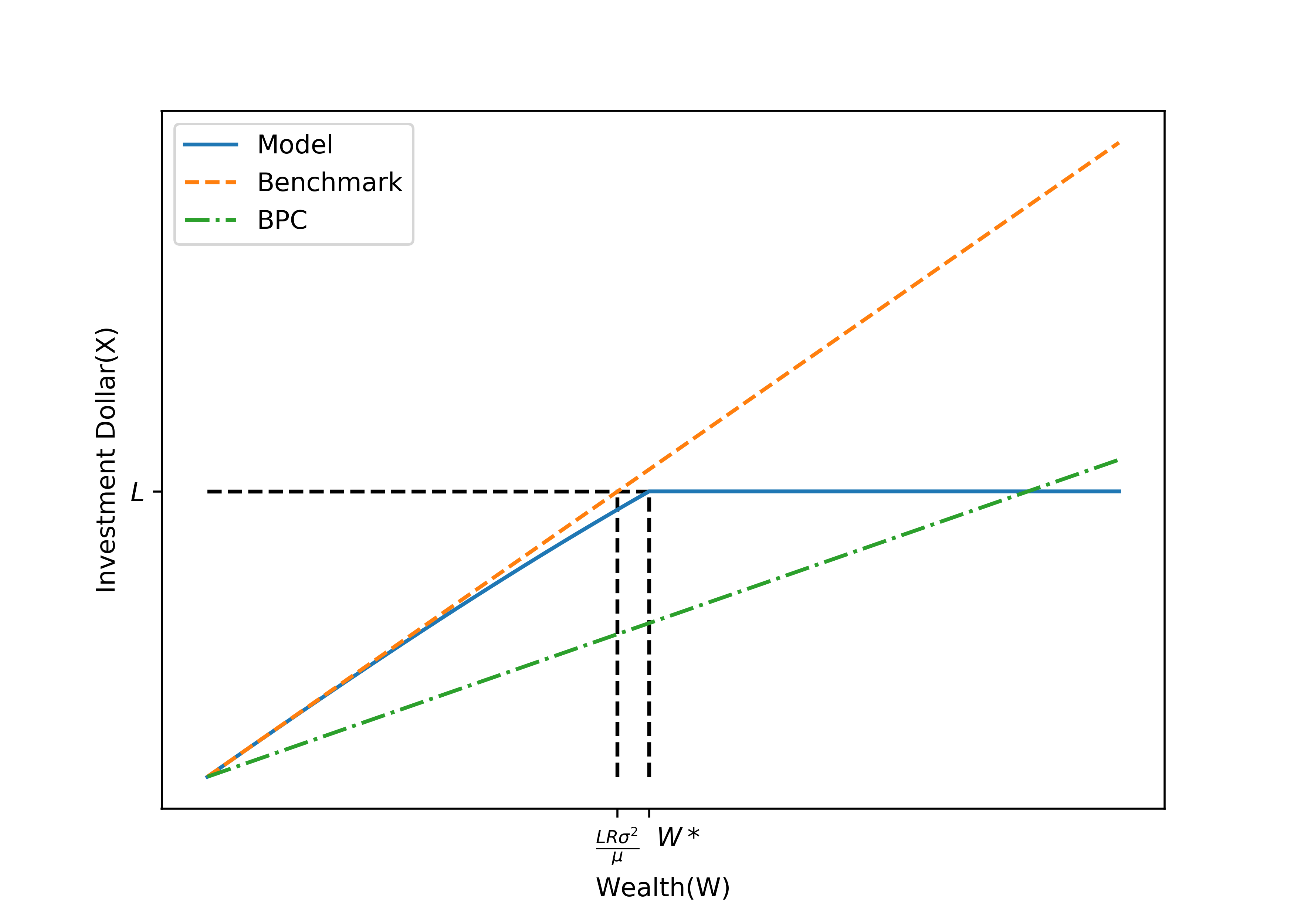}
\end{center}
					\caption{This figure displays the optimal portfolio strategy under three different strategies for $R > 1$.  ``Model" denotes the model in Problem (B) under a risk capacity constraint $X_t  \le L = \frac{0.7}{3}W_0$. Parameters are $\mu = 0.1, \sigma = 0.3, R = 1.5, l = 0.7/3.$ By calculation, the threshold level of the wealth is $W^* = 339,168$.  ``Benchmark" denotes the optimal dollar amount in Proposition \ref{pr:general without constraint} in the absence of the constraint on the risky asset investment. Finally, ``BPC" denotes the optimal strategy in Proposition \ref{prop:alternative} under a leverage constraint that $X_t \le \frac{1}{2} \frac{\mu}{R \sigma^2} W_t$.  }
\end{figure}

				\clearpage
\vspace{10pt}

\begin{figure}[!tp]
					\label{fig:Dollar2A}
\begin{center}
					\includegraphics[width=.98\textwidth] {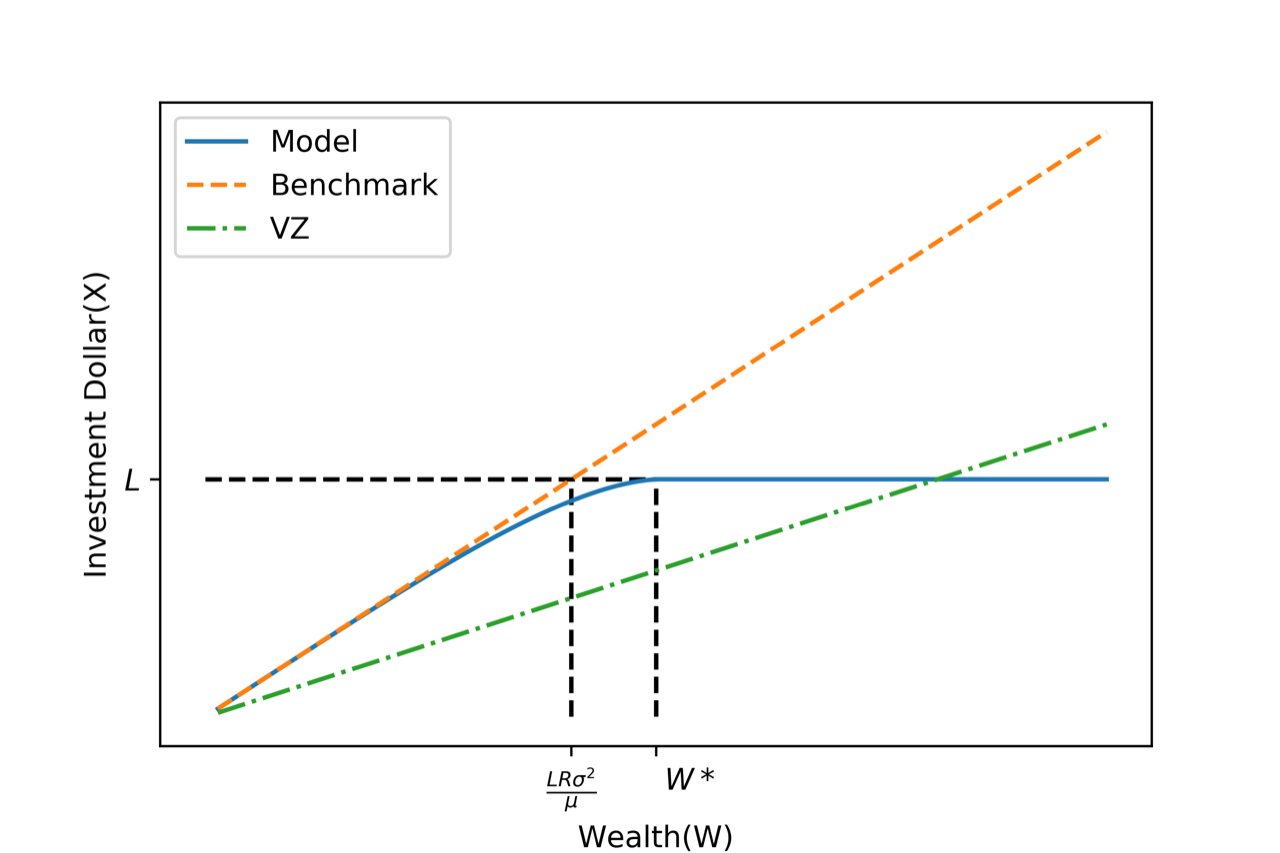}
\end{center}
					\caption{This figure displays the optimal portfolio strategy in Problem (A) under three different strategies.  ``Model" denotes the model in Problem (A) under a risk capacity constraint $X_t  \le L = 0.3 W_0$. Parameters are $\mu = 0.1, \sigma = 0.3, R = 1.5, l = 0.3.$ ``Benchmark" denotes the optimal dollar amount in Merton's model. Finally, ``VZ" denotes the optimal strategy under a leverage constraint that $X_t \le \frac{1}{2} \frac{\mu}{R \sigma^2} W_t$, which is solved in Vila and Zariphopoulou (1997). }
\end{figure}

%
%

				\clearpage
\vspace{10pt}
		
					\begin{figure}[!tp]
					\label{fig:Compare}
\begin{center}
					\includegraphics[width=.98\textwidth] {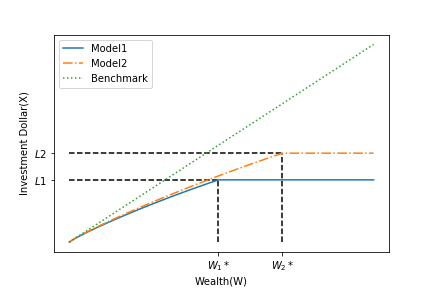}
\end{center}
					\caption{This figure displays the effect of the risk capacity level, $L$, on the investing strategy in Problem (B). The parameters are the same as in Figure 2. As shown, the higher the capacity level $L$, the higher the dollar amount in the risky asset. The figure also demonstrates that the threshold, $W^*$, positively depends on $L$.  The risk capacity level $L$ affects both the expected level of spending and the investing strategy even when the portfolio value is smaller than this threshold.
					}
					\end{figure}
\end{document}